\documentclass[twocolumn,trackchanges]{aastex7}

\newcommand\aastex{AAS\TeX}

\begin{document}

\title{Template \aastex v7 Article with Examples\footnote{Footnotes can be added to titles}}

 \def\gtorder{\mathrel{\raise.3ex\hbox{$>$}\mkern-14mu
    \lower0.6ex\hbox{$\sim$}}}
\def\ltorder{\mathrel{\raise.3ex\hbox{$<$}\mkern-14mu
    \lower0.6ex\hbox{$\sim$}}}

\title{The Impact of Seyfert Jets on Galaxy Evolution Across Major Scaling Relations}

\author[0009-0003-4860-8488]{Julianne Goddard}
\affiliation{Department of Physics and Astronomy, University of Kentucky, Lexington KY 40506-0055, USA}\email{julianne.goddard@uky.edu}

\author[0000-0002-1233-445X]{Isaac Shlosman}
\affiliation{Department of Physics and Astronomy, University of Kentucky, Lexington KY 40506-0055, USA}
\email{E-mail: isaac.shlosman@uky.edu}

\author[0000-0002-0071-3217]{Emilio Romano-Diaz}
\affiliation{Argelander-Institut f\"ur Astronomie, Universit\"at Bonn, Auf dem H\"ugel 71, 53121 Bonn, Germany}\email{E-mail: emiliord@uni-bonn.de}

\begin{abstract}
We analyze a suite of high-resolution cosmological zoom-in simulations of jetted Seyfert galaxies over $z\ltorder10$ projected on the major scaling relations, comparing trajectories of `normal' versus jet-hosting galaxies. Models include thermal and mechanical jet feedback launched from supermassive black holes (SMBHs) seeded at $z\sim9.1$ and $z\sim3.7$ with $M_\bullet\sim10^6\,M_\odot$ in galaxies within dark matter halos of ${\rm log}\,M_{\rm halo}/M_\odot\sim11.8$ at $z=0$. A single parameter, the SMBH accretion efficiency, has been varied resulting in $L_{\rm jet}\sim10^{40-42}\,{\rm erg\,s^{-1}}$, and SMBH accretion rates range between $\sim 0.2-10^{-4}$ of the Eddington rate. We find that jet feedback (1) suppresses central star formation rates (SFRs), redistributes gas to larger radii, (2) generates long-lived expanding shocks that couple to the ISM and CGM, (3) reduces stellar mass ($M_*$), shifting galaxies toward lower central concentrations, and (4) alters host trajectories on the $M_{\rm halo}-M_*$, specific SFR$-M_*$, $M_\bullet-\sigma_{\rm bulge}$, Mass$-$Metallicity, Kennicutt-Schmidt, and baryonic Tully-Fisher relation planes. Specifically, we find that jetted Seyferts live longer in the green valley and more frequently move to the quenched region in comparison to the non-jetted galaxies. Despite producing only transient quenching, Seyfert jets cause persistent structural, kinematic and chemical signatures, including flatter rotation curves, elevated CGM metallicities, and reduced cold gas clumping. (5) Early SMBH seeding and stronger jets amplify these effects, yielding galaxies that lie systematically closer to some of the empirical relations, e.g., $M_{\rm halo}-M_*$, while showing offsets for others, e.g., Kennicutt-Schmidt, and demonstrating that low-luminosity Seyfert jets can exert a significant long-term influence on galaxy evolution.   
\end{abstract}

\keywords{\uat{AGN host galaxies}{2017} --- \uat{Circumgalactic medium}{1879} --- \uat{Scaling relations}{2031} --- \uat{Hydrodynamical simulations}{767} --- \uat{Relativistic jets}{1390} ---  \uat{Seyfert galaxies}{1447}}


\section{Introduction} 
\label{intro}

Statistical analysis of large-scale galaxy properties is a powerful tool for studying galaxy evolution. A number of relationships between the basic parameters characterizing galaxies have been invoked to interpret how they grow and evolve. Among the most prominent of these are the correlations between the dark matter (DM) halo and stellar masses in galaxies, the $M_{\rm halo}- M_{*}$ relation \citep{behroozi19}, and between the supermassive black hole (SMBH) masses and velocity dispersion of galactic bulges, the $M_\bullet-\sigma_{\rm bulge}$ relation \citep{ferrarese00,gebhardt00}. Additional relations between various physical properties of galaxies have been defined and serve as benchmarks of star formation activity, most prominently the Kennicutt-Schmidt (K-S) relation \citep{schmidt59, kennicutt98}, the stellar mass - gas metallicity relation (MZR) \citep{lequeux79}, and structural scaling relations, e.g., Tully-Fisher \citep{tully77}, Faber-Jackson \citep{faber76}, the fundamental plane \citep{djorgovski87}, and more.
 
The intricacies of star formation (SF) provide the key to understanding galaxy evolution. There is a long list of external and internal factors that determine star formation rates (SFRs), these include: feedback from active galactic nuclei (AGN) and supernovae (SN), stellar and galactic winds, galaxy mergers and interactions, filamentary and diffuse accretion in DM halos, and the presence of stellar and gaseous bars. Together, these comprise the so-called baryon cycle \citep[e.g.,][]{keres05,oppenheimer06,shlosman13,sadoun16,sadoun19,tumlinson17a,tumlinson17b,kewley19}. AGN feedback includes radiative, mechanical, and thermal feedback from SMBH accretion disks and relativistic jets, and injection of high energy particles in the form of cosmic rays. These can affect the dynamic and thermodynamic properties of the gas distribution and, in particular, the star-forming gas in galaxies. 

Powerful jets in quasars have been studied for decades, both theoretically, numerically, and observationally \citep[e.g.,][and refs. therein]{bower06,croton06,fabian12}. Their less energetic counterparts in Seyfert galaxies are much less known, partially because their detection is more difficult, and their properties are based mostly on the local population of Seyferts. Seyfert AGN are characterized by lower-mass SMBHs, $\leq 10^8 \,M_\odot$, and lower-luminosities, $< 10^{46}\,{\rm erg\,s^{-1}}$. Seyfert jets inject energy from parsec scales up to the circumgalactic medium (CGM) and beyond, in a highly collimated, anisotropic fashion. This can affect a wide range of scales, from the SMBH radius of influence to the interstellar medium (ISM) and CGM, and affect the cooling flows from cosmological filaments. 

AGN in both the low- and high- redshift Universe are typically identified through one of several well-established observational signatures. For example, optical emission-line ratios (e.g., BPT diagrams, \citet{baldwin81}), mid-IR color selections tracing hot dust \citep[e.g.,][]{stern05}, X-ray luminosity or hardness ratios \citep[e.g.,][]{brandt05}, radio continuum emission \citep[e.g.,][]{condon91}, and broad emission-line widths \citep[e.g.,][]{osterbrock89} --- are all common AGN identification methods. While these diagnostics successfully identify luminous or unobscured AGN, each suffers from selection biases that are particularly significant for low-luminosity jets. Optical line-ratio methods can become unreliable at high redshift due to evolving ISM conditions and strong star formation \citep[e.g.,][]{goulding09, kewley13}. X-ray and mid-IR selection can miss AGN that are weakly accreting, heavily obscured, or dominated by star formation \citep[e.g.,][]{assef13,donley12}. Furthermore, radio surveys often lack the sensitivity or spatial resolution to detect parsec-scale or intermittently active jets \citep[e.g.,][]{radcliffe18}. As a result, many galaxies experiencing feedback may be classified as “normal” starforming systems, despite hosting energetically important AGN-driven jets. 

With the arrival of the latest generation of space and ground-based telescopes, we are granted the first spectroscopic studies of moderate and lower-luminosity AGN, up to $z\sim 7$ \citep{maiolino24, pacucci23}. Rough estimates for morphology, mass, SFRs, SMBH masses, and even mechanical luminosity of jets have been obtained.  This new generation of observations has opened a window into the high-$z$ evolution of many galaxy scaling relations, including the stellar mass-size relation ($M_*-R_{\rm gal}$), UV luminosity–mass ($L_{\rm UV}-M_*$) relation, and rest-frame optical/near-IR color-magnitude ($[g-H]-M_{\rm H}$) diagrams for galaxies at $z\sim 0.5-7$. 

Simulations can extend our knowledge of how small-scale jets affect their host galaxies across these scaling relations. Large-scale cosmological simulations, such as EAGLE \citep{schaye15}, IllustrisTNG \citep{nelson19}, and SIMBA \citep{dave19}, are largely designed to reproduce the observed relationships between galaxy properties. These simulations show reasonable qualitative agreement with observations. For example, the normalization and slope of the star-forming main sequence decline from high to low redshifts \citep{furlong15, donnari19}, the Mass-Metallicity relation (MZR) becomes more metal-rich over time \citep{torrey19,zenocratti22}, and galaxies approach the $M_{\rm halo}-M_{*}$ relation when sufficient feedback is introduced \citep{schaye15,nelson18}. However, because most large-volume simulations suffer from computational constraints, they cannot resolve the details of direct interactions between the feedback and ISM/CGM. As a result, they provide statistical baselines but often lack the ability to interpret detailed observational measurements, especially in systems where small-scale processes, like jets in Seyferts, dominate the evolution.

This is precisely where targeted, high-resolution zoom-in simulations provide an advantage. By re-simulating the same galaxy under systematically varied jet-feedback strengths, we can isolate how AGN jet power shapes the evolution of Milky-way mass galaxies on the scaling relations and offer a causal interpretation that cannot be extracted from statistical simulation suites alone. Observational studies detect jet-driven disturbances and gather statistical analysis of AGN host systems, but they cannot follow the multi-Gyr dynamical, thermodynamic, and chemical evolution that determine where a galaxy lands on the scaling relations or how it will evolve in the future. Thus, our goal is to supply a physically interpretable, time-resolved perspective, bridging the gap between large-volume statistical modeling and observations.

Previous zoom-in numerical works, especially those involving FIRE-2 and 3 \citep[][]{wellons23,byrne24}, have studied the effects of various feedback recipes on the properties of galaxies in different mass ranges, but largely provided only the final results at $z=0$, or terminated evolution at $z=0.5$, and do not look into the isolated effect of jet feedback. A number of other works have studied feedback in Seyfert galaxies in a cosmological context \citep[e.g.,][]{okamoto08,irodotou22}, but did not model collimated energy deposition by jets. 

In this work, we analyze the evolution of jetted Seyfert galaxies on the major scaling relations across cosmic time, using a suite of high-resolution zoom-in cosmological simulations. Our simulations implement AGN feedback in the form of collimated jets. In \citet[hereafter Paper\,I]{goddard25a}, we presented the effects of the jet-feedback on galaxies in their final state, at $z=0$, by varying the SMBH accretion efficiency and observing the changes in morphology and thermodynamic state of the galaxy.  In \citet[hereafter Paper\,II]{goddard25b},  we dealt with the evolution of our simulations for $z\ltorder 10$, and focused on how accretion efficiency and SMBH seeding time influence the evolution of the structure of the galaxy and its environment. Here we study the effect of jet feedback on the galaxy’s evolution across several key scaling relations: $M_{\rm halo}-M_*$, specific star formation rate (sSFR) - $M_*$, $M_{\bullet}-\sigma_{\rm bulge}$, MZR, K-S, and baryonic Tully-Fisher (BTFR) relations.  By placing our results in the context of existing observational and theoretical work, we aim to clarify the role of the Seyfert jet feedback on the evolution of host galaxies and their immediate environment.

The structure of this paper is as follows. Section\,\ref{sec:num} describes the numerical implementation, section \ref{sec:results} presents our results, Section\,\ref{sec:discussion} discusses these results and places them in context with existing and future numerical and observational studies.  

\section{Numerics} 
\label{sec:num}

A suite of high-resolution cosmological zoom-in simulations have been performed using the $N$-body/hydro code \textsc{gizmo}, using the MFM hydro solver \citep{hopkins15}. The full technical details of numerical setup have been given in Papers\,I and II. Only the most necessary details are given here. The initial conditions (ICs) were generated at $z=99$ using the \textsc{music} code \citep{hahn11} within a box of $50\,h^{-1}$\,Mpc. They use the \citet{planck16} $\Lambda$CDM concordant model, with $\Omega_{\textrm m} = 0.308$, $\Omega_\Lambda = 0.692$, $\Omega_{\textrm b} = 0.048$, $\sigma_8 = 0.82$, and $n_{\textrm s} = 0.97$, with the Hubble constant $h = 0.678$ in units of $100\,{\textrm {km}\,\textrm s^{-1}\,\textrm {Mpc}^{-1}}$.

From the parent, uni-grid, DM-only simulation halos have been identified by the group finder \textsc{rockstar} \citep{behroozi12}, with a Friends-of-Friends (\textsc{FoF}) linking length of $b=0.28$. A single DM halo was chosen with virial radius, $R_{\textrm {vir}}$, and virial mass, $M_{\textrm {vir}}$  defined in terms of $R_{200}$ and $M_{200}$ \citep[e.g.,][]{navarro96}, where $R_{200}$ is the radius within which the mean interior density is 200 times the critical density of the universe at that time, and $M_{200}$ is the corresponding enclosed mass. The $M_{\textrm {vir}}$ and $R_{\textrm {vir}}$ values refer to the DM component {\it only}. In the presence of baryons, we define $M_{\rm halo}$ as the total baryonic$+$DM mass inside R$_{\rm vir}$.

In order to focus on the effect of the accretion efficiency, $\epsilon$, we have used a set of models, evolved from identical initial conditions, and varied only this parameter. We have run two suites of four simulations each that differ only in $\epsilon$. Because all models use the same DM halo, they are embedded in the same large-scale environment, allowing for a direct comparison between the jet models. Of course, jet effects may be influenced by specific evolutionary events unique to this halo. Modeling a population of different halos while holding the model parameters fixed constitutes a different approach which is outside the scope of this work. The evolutionary epochs and environmental influences observed for this halo are discussed further in section\,\ref{sec:time}.

The effective number of particles (DM and baryons) in our simulations is $2\times 4,096^3$, resulting in mass resolution per particle of $3.6\times 10^4\,{\rm M_\odot}$ for the gas and stars, and $1.9\times 10^5\,{\rm M_\odot}$ for the DM. The minimal adaptive gravitational softening in co-moving coordinates for the gas is 1\,pc, for stars 20\,pc and 200\,pc for DM.

\begin{deluxetable*}{ccccccccccccccc}[ht!]
\tabletypesize{}
\tablecolumns{14}
\tablecaption{Model Properties at $z=0$\label{tab:galprops}}
\tablehead{
\colhead{Model} & \colhead{log\,$M_{\rm halo}$} & \colhead{log\,$M_*$} & \colhead{log\,$M_\bullet$} & \colhead{$f_{\rm gas}$} &\colhead{log SFR} & \colhead{log $Z_{\rm gas}$} & \colhead{log $Z_*$} & \colhead{$R_{\rm 1/2}$} & \colhead{$R_{\rm e}$} & \colhead{log\,$L_{\rm bol}$} & \colhead{$M_{*}/L_{\rm bol}$} &\colhead{log\,$L_{\rm jet}$} \\
\colhead{} & \colhead{M$_\odot$} & \colhead{M$_\odot$} & \colhead{M$_\odot$} & \colhead{} &\colhead{$M_{\odot}\,{\rm yr^{-1}}$} & \colhead{$Z_\odot$} & \colhead{$Z_\odot$} & \colhead{kpc} & \colhead{kpc} & \colhead{L$_\odot$} &  & \colhead{erg s$^{\rm -1}$} }
\startdata {$\epsilon_{\rm 0, LBH}$} & 11.92 & 10.97 & --- & 0.09 & 0.32 & -0.80 & 0.04 & 1.3 & 2.2 & 10.51 & 2.9 & --- \\ 
{$\epsilon_{\rm 5, LBH}$} & 11.91 &  10.94 & 7.73 & 0.07 & -0.48 & -0.98 & 0.04 & 1.5 & 1.8 & 10.15 & 6.2 & 41 \\
{$\epsilon_{\rm 15, LBH}$} & 11.88 & 10.68 & 7.86 & 0.13 & -0.92 & -0.75 & -0.01 & 1.6 & 2.4 & 9.88 & 6.3 & 41 \\
{$\epsilon_{\rm 50, LBH}$} & 11.87 & 10.58 & 7.86 & 0.05 & -2.66 & -0.83 & -0.02 & 1.6 & 1.5 & 9.62 & 9.1 & 40 \\
\hline
{$\epsilon_{\rm 0, EBH}$} & 11.92 & 11.00 & --- & 0.11 & 0.09 & -0.42 & 0.11 & 1.7 & 3.5 & 10.42 & 3.8 & --- \\ 
{$\epsilon_{\rm 5, EBH}$} & 11.90 &  10.77 & 7.11 & 0.13 & -0.03 & -0.10 & 0.07 & 2.6 & 5.6 & 10.32 & 2.8 & 40 \\
{$\epsilon_{\rm 15, EBH}$} & 11.89 & 10.72 & 7.98 & 0.15 & -0.27 & -0.16 & 0.04 & 2.3 & 6.2 & 10.06 & 4.6 & 41 \\
{$\epsilon_{\rm 50, EBH}$} & 11.86 & 10.64 & 8.19 & 0.06 & -1.14 & -0.10 & 0.04 & 2.0 & 2.5 & 9.79 & 7.1 & 42 \\
\hline
\enddata
\tablecomments{Columns: (1) model name; (2) DM + baryonic mass inside $R_{\rm vir}$; (3) stellar mass; (4) SMBH mass (5) gas fraction $f_{\rm gas}=M_{\rm gas}/(M_{\rm gas}+M_{*})$; (6) SFR averaged over the final 30\,Myr; (7) gas metallicity; (8) stellar metallicity; (9) stellar half-mass radius; (10) stellar half-light radius; (11) bolometric galaxy luminosity; (12) $M_*/L_{\rm bol}$; (13) total jet luminosity, $L_{\rm jet}=(1/2)\eta\dot M_\bullet(v_{\rm jet}^2 +3 kT/m_{\rm p})$.  NOTE: All galaxy properties are defined inside $0.1R_{\rm vir}\sim 23$\,kpc. Column (10) is found using mock-photometry as described in section\,\ref{sec:num} using the SDSS R-band.}
\end{deluxetable*}

The group-finding algorithm \textsc{hop} \citep{eisenstein98} has been used to identify galaxies, using the outer boundary threshold of baryonic density of $10^{-4}\,n^{\textrm {SF}}_{\textrm {crit}} = 10^{-2}\,{\rm cm^{-3}}$, which ensured that both the host SF gas and the lower density non-SF gas are roughly bound to the galaxy \citep{romano-diaz14}. This assures that identified galaxies are not imposed with a particular geometry. Following Papers\,I and II, the \textsc{hop}-defined galaxy size was found to be similar to 0.1$R_{\rm vir}$, which is used here.

Stars form only where gas is self-gravitating, namely, when
\begin{equation}
        (\nabla v)^2+2c_\mathrm{s}^2 < 8 \pi G \rho\,,
\end{equation}
where $c_{\rm s}$ is the sound speed in the gas (divided by the effective particle size), $\rho$ is the gas density, and $\nabla v$ is the physical velocity gradient. To determine self-gravity, we rely on the virial parameter, $\alpha_\mathrm{vir}$, defined as 
\begin{equation}
    \alpha_\mathrm{vir}=[(\nabla v)^2+2c_\mathrm{s}^2]/8 \pi G \rho\,.
\label{eq:avir}
\end{equation}
The self-gravity condition is met when $\alpha_\mathrm{vir} < 1$  (\cite{hopkins13}). The star formation efficiency (SFE) is calculated from the virial parameter following the model by \cite{padoan12}:
\begin{equation}
    SFE= \mathrm{exp}(-1.4\alpha_\mathrm{vir}^{1/2}) ~.
\label{eq:sfe}
\end{equation}
Once the SFE is determined, stars form stochastically (\cite{springel03}).  Each star particle represents an entire population of stars with a mass distribution following the \cite{chabrier03} Initial Mass Function (IMF).

Gas heating and cooling from $10^{10}$\,K down to 10\,K are implemented, including H and He ionization$+$recombination, collisional, free-free, dust collisional, cosmic ray, and Compton effects, as well as metal-line \citep{wiersma09}, fine-structure, and molecular cooling \citep{hopkins18,hopkins22}. Metal enrichment is included: the metallicity increases in the star-forming gas and scales with the fraction of stars that turn into SN, and the metal yield per SN (see below). A total of 11 metal species were followed in both in gas and stars, including H, He, C, N, O, Ne, Mg, Si, S, Ca, and Fe. The H$_2$ abundances used for cooling calculations are estimated from the analytic fitting function of \citet{krumholz11}.   

Very little is understood about how SMBHs grow in the early Universe, thus we vary the seeding time of the SMBHs between two suites to study the effect that SMBH seeding time has on the evolution of the host galaxy. In the early-seeded (EBH) models the SMBH is seeded when the stellar mass of the galaxy reaches $10^{8}\,M_\odot$, this occurs at $z\sim 9.1$. The late-seeded (LBH) models seed SMBHs when the halo mass reaches $10^{11}\,M_\odot$, which happens at $z\sim 3.7$. The SMBHs are seeded only in the central halo in order to examine the influence of a single AGN on its host and its environment. 
SMBHs are initialized with $M_\bullet\simeq10^6\,M_\odot$, and grow by accreting surrounding gas, with an accretion rate, $\dot M_{\rm grav}$, determined based on gravitational torques \citep{shlosman89}, and calculated using the \citet{hopkins11} method. The spin axis of the SMBH evolves by inheriting angular momentum from the gas that it accretes.

\begin{figure}[ht!]
\center
\includegraphics[width=0.8\linewidth]{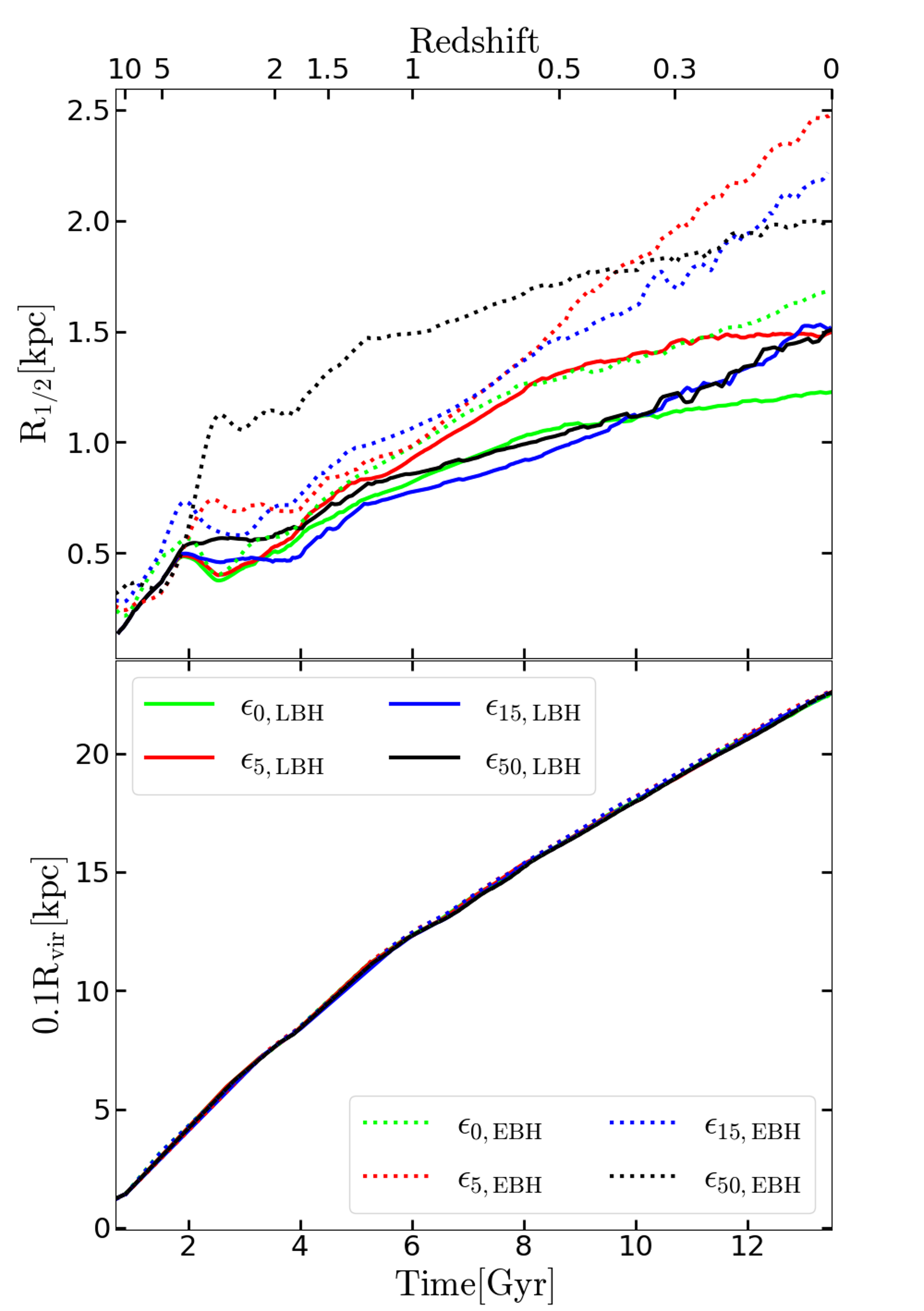}
\caption{Comparison of the galaxy radius evolution defined using 10\% the halo virial radius, 0.1R$_{\rm vir}$ (bottom),  versus the stellar half-mass radius, $R_{\rm 1/2}$ (top), for our eight EBH and LBH models. There are individual lines representing 0.1R$_{\rm vir}$ for each model, but they all sit together, as variations in R$_{\rm vir}$ are minimal between the models. Note that the final value of $0.1R_{\rm vir}$ is $\sim 10-20\times$ that of $R_{\rm 1/2}$.
\label{fig:Rcomp}}
\end{figure}

The tentative accretion rate, $\dot M_{\rm grav}$, has been multiplied by an efficiency parameter, $\epsilon$ \citep{angles-alcazar17}, to capture the effects of unresolved processes affecting gas inflow. The sole parameter varied between the AGN models was $\epsilon$. We set it to $\epsilon=0\%$, 5\%, 15\%, and 50\%, and denote these models as $\epsilon_0$, $\epsilon_5$, $\epsilon_{15}$ and $\epsilon_{50}$, respectively.  The final accretion rate is
\begin{equation}
    \dot{M}_\bullet=\epsilon\dot{M}_{\rm grav}.    
\end{equation}
The SMBH accretion rate measured as a fraction of the Eddington accretion rate, $\dot M_{\rm Edd}$, is 
\begin{equation}
    f_{\rm Edd}=\dot M_\bullet/\dot M_{\rm Edd},
\end{equation}
where $\dot M_{\rm Edd}=L_{\rm Edd}/c^2$, and $L_{\rm Edd}$ is the Eddington limit for $M_\bullet$. 

To model the AGN jets, we use hyper-refined particle spawning \citep{torrey20, su21}. Particles of mass $10^3\,M_\odot$ are spawned at a fixed fraction\footnote{Taken as the minimum of 0.125\,SMBH kernel radius or half the distance between the SMBH and the closest gas particles.} of the SMBH kernel radius with an initial velocity of $3\times 10^4\,{\rm km\,s^{-1}}$ and a temperature of $10^{10}$\,K. Jet particles are launched along the spin axis of the SMBH with a mass loading $\eta$=0.1, meaning 10\% of the accreted gas is returned in the form of jet particles. The launching rate is thus $\dot{M}_{\rm jet}=\eta\dot{M}_\bullet$, giving a jet energy injection rate of 
\begin{equation}
    L_{\rm jet}=(1/2)\eta\dot M_\bullet(v_{\rm jet}^2 +3 kT/m_{\rm p}),
\label{eq:Ljet}
\end{equation}
where $v_{\rm jet}$ is the jet particle velocity, and $T$ is their temperature. Eq.\,\ref{eq:Ljet} describes both kinetic and internal energies of the jet particles. The velocities of the spawned particles are initially perfectly collimated. Once created, the jet particles interact hydrodynamically in the same way as any of the other gas particles. When they decelerate to 25\% of their initial velocity, and enter the kernel radius of another gas particle in a head-on trajectory, the jet particles re-merge and the mass-weighted properties of the two particles are averaged.    

We have invoked the  mechanical feedback from SN-type\,II with the numerical prescription given in \citet[][see also \citet{goddard25a}]{hopkins18}.  In the LBH models, each SN event injects an energy of $1\times 10^{51}$\,erg, a mass of 14.8\,$M_\odot$, and a metal mass of 2.6\,$M_\odot$ into the surrounding gas within a radius of 200\,pc. For EBH models, each SN event injects an energy of $5\times 10^{51}$\,erg within 750\,pc. 

The motivation for adjusting the SN feedback prescription in the EBH models comes from a previous study \citep[][]{roca21}, which argues in favor of enhanced SN feedback based on the comparison of a number of numerical codes.  As this change to the stellar feedback routine introduces another parameter to the comparison between our EBH and LBH models, we have run a {\it test} model to quantify this comparison. This test model uses the original stellar feedback prescription from the LBH models, but seeds the SMBH using the EBH model criteria. Description of this test model, $\epsilon_\mathrm{5, EBH}$(test), and the information it provides towards disentangling the specific effects of the SMBH seeding time versus SN feedback prescription are discussed in detail in Paper\,II. Table \ref{tab:galprops} provides a summary of our models along with their final galaxy and halo properties.

We have used the 3-D radiation transfer code SKIRT \citep{baes03} to obtain post-processed mock-SEDs (spectral energy distributions) and photometry for our galaxies at $z=0$. Within SKIRT each stellar particle is treated as a stellar population with an initial mass function (IMF) as given in \citet{chabrier03} and an SED family \citep[e.g.,][]{bruzual93}. We assume dust content to be 20\% of the gas metallicity and we include secondary emission from the dust and iterations for dust self-absorption. For this analysis we include all gas and stars within $0.1R_{\rm vir}$, i.e., within the defined galaxy, and look only on the SDSS R-band to find the galaxy half-light radius, $R_{\rm e}$. The bolometric luminosity, $L_{\rm bol}$, was determined from the face-on projected flux integrated between $\sim 0.15-400\,\mu{\rm m}$. 

To compare simulated galaxies to observations is a challenge, and we explore the extent to which the definition of a modeled galaxy alters its evolutionary track superposed on scaling relations. Methods used to determine properties such as stellar mass, SFR, $Z_{\rm gas}$, and many other observables can differ profoundly. Hence, one needs to be careful exploring the differences between modeled and observed galaxies, whether they originate from the methods applied or are real. Numerous examples exist in the literature where altering definitions or taking extra steps to `mock' observations have resulted in significant changes to the measured properties \citep[e.g.,][]{torrey19,cochrane23}. 

We define three characteristic radii to be used in our analysis, namely, (1) the galaxy size, $0.1R_{\rm vir}$, (2) the galaxy stellar half-mass radius, $R_{1/2}$, which is defined by the radius of a sphere enclosing half of the total stellar mass, and (3) the half-light radius, $R_{\rm e}$, determined using mock-photometry, as described above.  

Observed galaxies are most commonly defined by their half-light radii, Petrosian radius \citep{petrosian76}, or similarly. Thus, we expect values related to the photometry to provide the most consistent results between theory and observations. Due to the time constraint with running SKIRT, we only present the half-light radii values at $z=0$ in Table\,\ref{tab:galprops}. At the final redshift, $R_{\rm e}$ is located between $R_{1/2}$ and $0.1R_{\rm vir}$, but significantly closer to the former.  For this reason, we chose to include the evolution of $R_{\rm 1/2}$ for all models, together with $0.1R_{\rm vir}$. The time evolution of these is shown in Figure\,\ref{fig:Rcomp}. Note that $0.1R_{\rm vir}$ evolution is approximately linear with time for $z\gtorder 2$, changes its slope between $z\sim 2-1$, and then continues linearly with a smaller slope. The half-mass radii for the EBH AGN models separate from LBH models and stay above them for $z\ltorder 2$. 

Additional technical details regarding the simulation setup, the SF and SN feedback, as well as the seeding of the SMBH and its mechanical feedback in the form of jets, have been outlined in Papers\,I and II.

\begin{figure}[ht!]
\center
\includegraphics[width=0.9\linewidth]{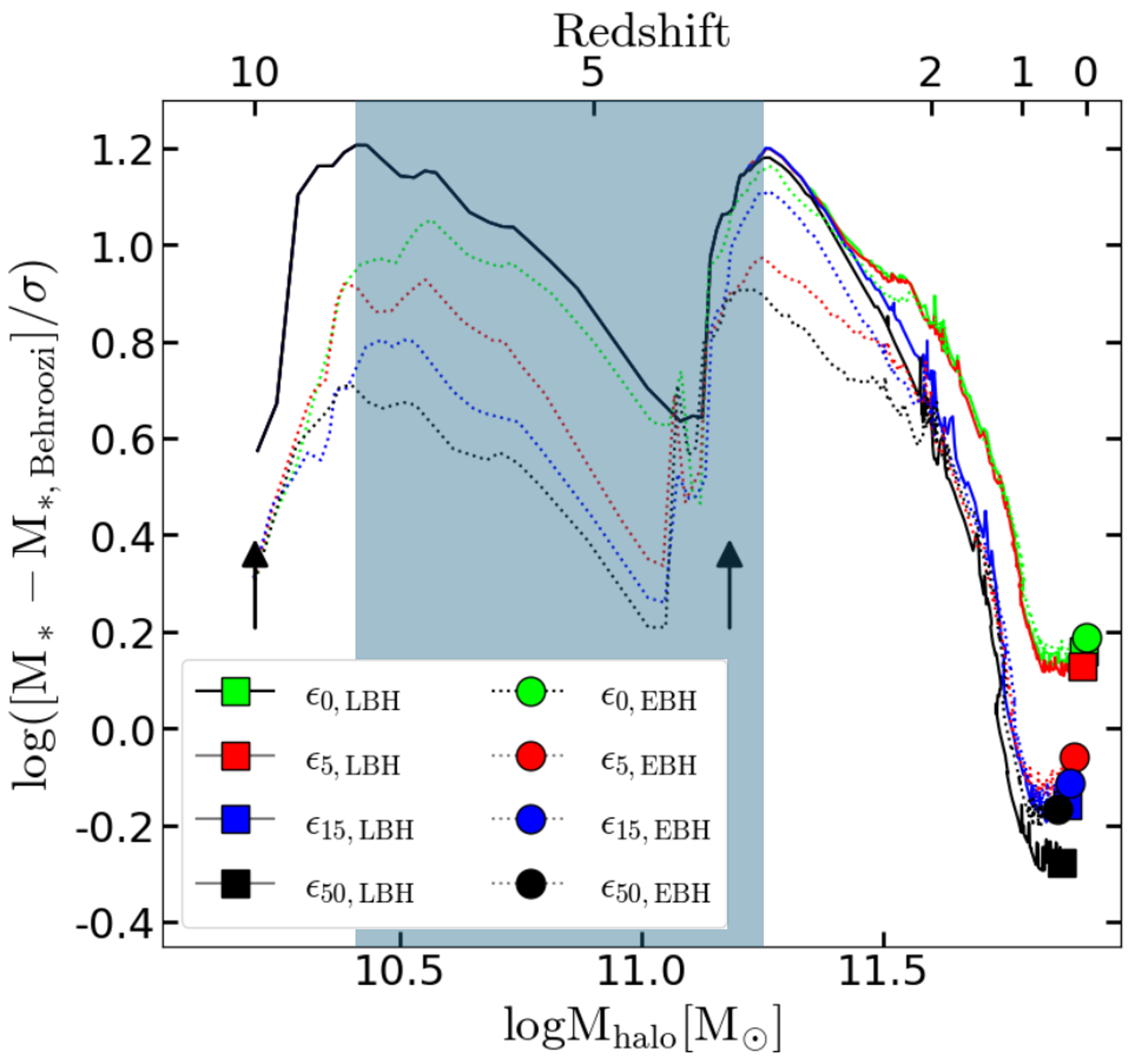}
\caption{Evolutionary tracks of modeled galaxies on the $M_{\rm halo}-M_*$ plane. The $y$-axis shows $M_*-M_{\rm Behroozi}$ --- the median relation from \citet{behroozi19} has been subtracted from our stellar masses, and this has been normalized by the scatter band of $\sigma=\pm 0.5$\,dex. This provides the distance of our galaxies from the median scaling relation as a function of $M_{\rm halo}$.  The shaded region indicates the time interval of the last major merger, and the black arrows show the seeding times of the SMBHs for the late-seeded (LBH) and early-seeded (EBH) models. Note that the growth of $M_{\rm halo}$ is very similar for all eight models, as such the top x-axis showing redshift remains approximately accurate for all simulated halos.
\label{fig:smbh}}
\end{figure}

\section{Results} 
\label{sec:results}

In this section we analyze the evolution of simulated Seyfert-type host galaxies with respect to several normal galaxy scaling relations over cosmological timescales, from $z\sim 9$ to $z=0$. Note that the $\epsilon_{0}$ models in both suites do not host an SMBH and serve as our benchmark, against which we can compare the effects of the AGN feedback. See Section \ref{sec:num} and Table \ref{tab:galprops} for more details on the simulation setup and galaxy properties. 

\subsection{Scaling relations involving $M_{\rm halo}$, $M_*$, and $M_\bullet$} 
\label{sec:figure2}

\begin{figure*}[ht!]
\center
\includegraphics[width=0.8\linewidth]{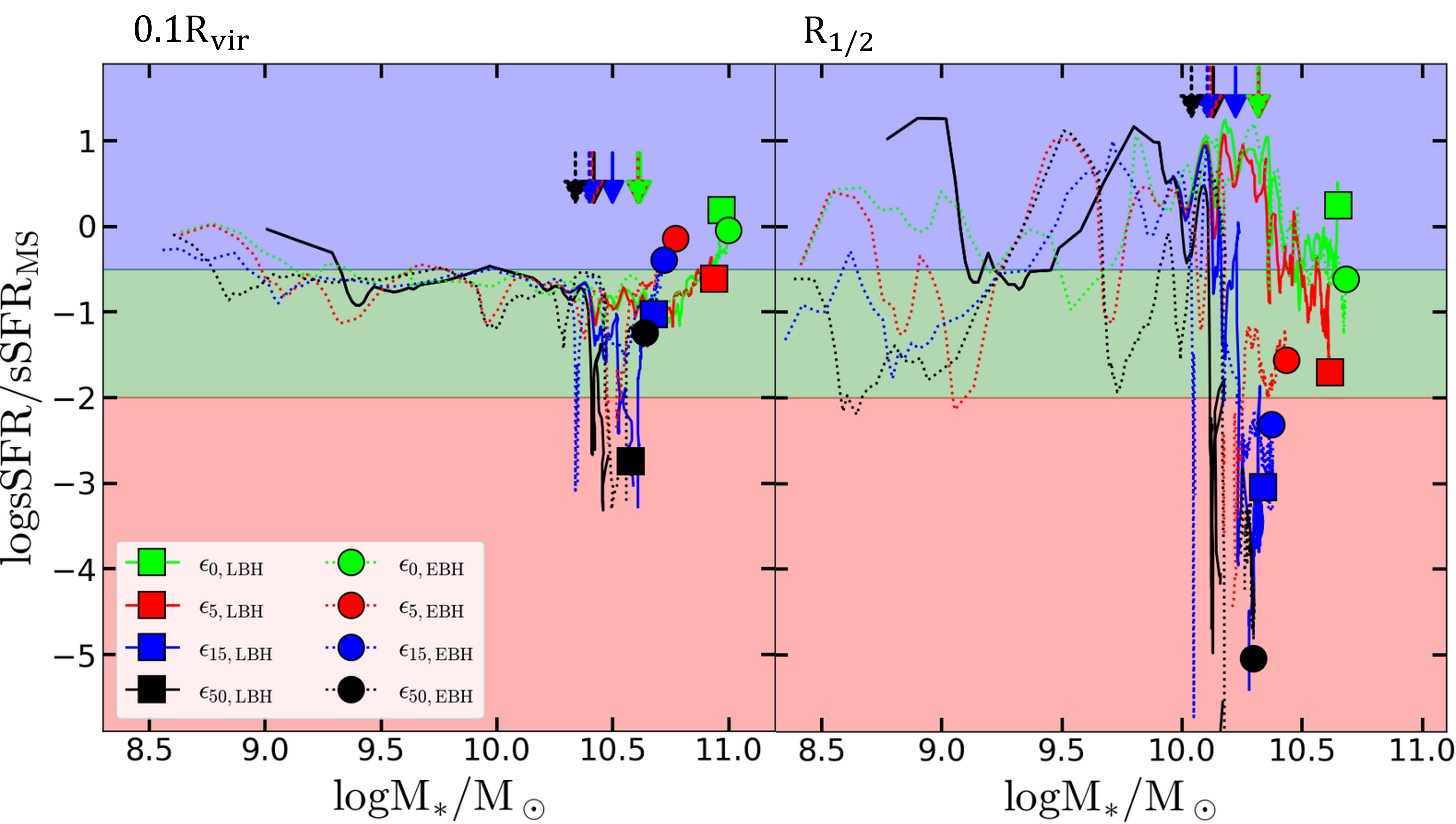}
\caption{Evolutionary tracks of modeled galaxies on specific SFR --- $M_*$ plane, normalized by the star-forming galaxy main sequence fits derived in \citet{Popesso23}. The colors indicate divisions between the star-forming (blue), green valley (green), and quiescent (red) regions. The left frame shows sSFR and $M_*$ inside $0.1R_{\rm vir}$ and the right represents both quantities inside $R_{1/2}$. The arrows show $z=2$ for each model, with the arrow color corresponding to the model marker color indicated in the legend. Note that the $\epsilon_{\rm 50,LBH}$ model within $R_{1/2}$ does not recover after the dip and stays below the Figure for the rest of the simulation.
\label{fig:sfms}}
\end{figure*}

\begin{figure}[ht!]
\center
\includegraphics[width=0.8\linewidth]{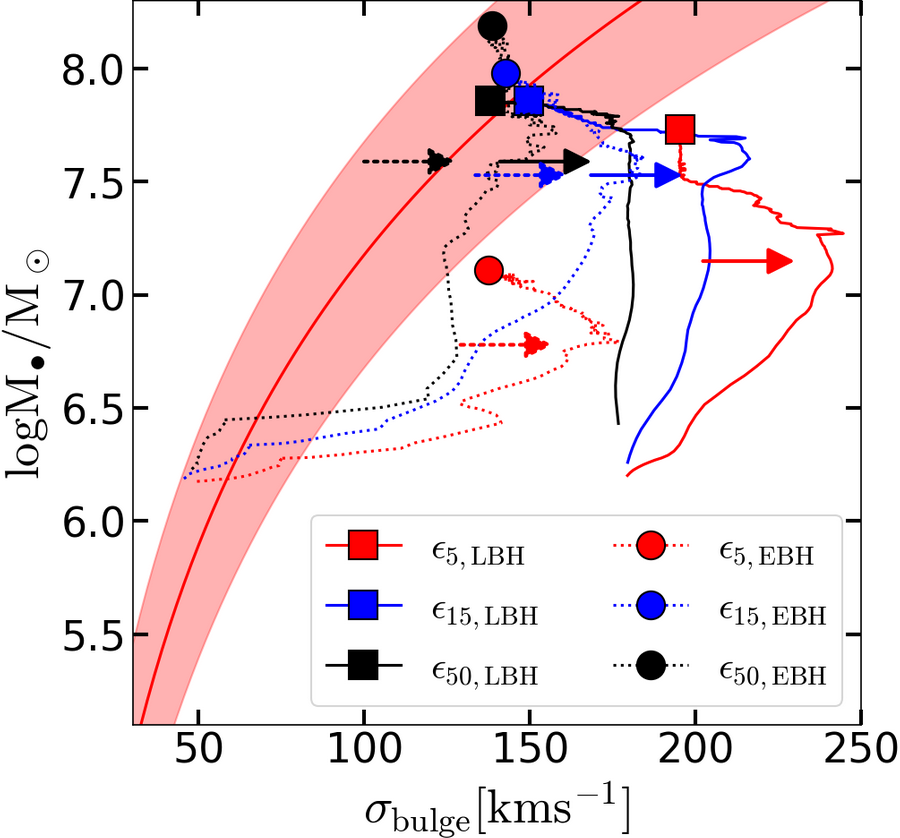}
\caption{Evolutionary tracks of modeled galaxies on the $M_{\bullet}$ - $\sigma_{\rm bulge}$  plane. The solid red line displays the median relation fit from \citet{kormendy13}, and the red shaded area shows 0.5\,dex scatter. The arrows indicate $z=2$ for each model, with the arrow color corresponding to the model marker color in the legend. Evolution starts at the seeding time of the SMBHs: $z\sim 9.1$ for the EBH models and $z\sim 3.7$ for the LBH models.  
\label{fig:Msigma}}
\end{figure}

We start with three fundamental scaling relations $M_{\rm halo}-M_*$ (Fig.\,\ref{fig:smbh}), sSFR --- $M_*$ (Fig.\,\ref{fig:sfms}), and $M_\bullet - \sigma_{\rm bulge}$ (Fig.\,\ref{fig:Msigma}), where $M_{\rm halo}$ is the total halo mass, sSFR is the specific SFR, and $\sigma_{\rm bulge}$ is the stellar velocity dispersion in the bulge. 

In Figure\,\ref{fig:smbh},\, we show the redshift evolution of the distance of our models from the median $M_{\rm halo}-M_*$ \citet{behroozi19} relation normalized by $\sigma = 0.5$\,dex error. The tracks have been annotated for each $\epsilon$. The final values at $z=0$ for each of the models are shown by the position of their associated marker.  

Figure\,\ref{fig:smbh} exhibits a number of interesting aspects of the modeled galaxy evolution on the $M_{\rm halo}-M_*$ plane in comparison to its median. Before a major merger at z$\sim$4.5, all models move away from the median, although the stronger feedback models always remain closer to the median. Note that a major merger also occurs around z$\sim$10, and the evolution shown before the shaded region reflects post merger starburst and stellar merging activity from this prior event. The period of the second major merger, highlighted by the shaded region, is characterized by sharp changes in the evolution on this plane, and can be divided into the DM halo merger, with models moving towards the median (i.e., the descending curve) and the subsequent galaxy merger, which is characterized by intensive SF (i.e., steeply ascending curve). The merging process takes about $\sim 1$\,Gyr, which corresponds to the distance between the two peaks and to the shaded region in the Figure.  

After the last major merger, all of the evolution tracks move towards the median, but the end products at $z=0$ are clearly segregated along the $y$-axis. Namely, the $\epsilon_0$ models remain outside the 0.5\,dex band, while the jetted galaxies lie within this band and stronger feedback brings them closer to the median curve. The steepest descent of the evolutionary tracks in Figure\,\ref{fig:smbh} occurs after the major merger, i.e., between $z\sim 3.5$ and $z\sim 1$, an about 5\,Gyr period characterized by intermediate and numerous minor mergers. After this time, all the tracks converge to horizontal. 

A number of corollaries can be elucidated from this evolution. Stronger SN feedback, early SMBH seeding time in the EBH models, and difference in the feedback strength measured by $\epsilon$, result in $M_*$ differing by a factor of $\sim 3$ at the end. Most remarkably, both LBH and EBH models with higher $\epsilon$ lie closer to the median at $z=0$.

Note that our Figure\,\ref{fig:smbh} has $M_{\rm halo}$ on the $x$-axis, which is the total of the baryonic and DM halo masses, while the Behroozi relation has the DM only mass on this axis. We have replotted this Figure replacing $M_{\rm halo}$ with $M_{\rm DM}$,  and found that it shifts the tracks very slightly up and to the left. However, all our conclusions remain valid.

Both frames of Figure\,\ref{fig:sfms} deal with evolution on the specific SFR --- $M_*$ plane. For each data point here, we have averaged the SFR over 30\,Myr, normalized it by $M_*$ at that time, and normalized by the main sequence SFR \citep[e.g.,][]{Popesso23}. Furthermore, we have divided the $y$-axis into three regions corresponding to star-forming (blue), green valley (green), and quenched SF (red).  The 'star-forming' region applies to galaxies sitting above the line drawn at 0.5\,dex below the star-forming main sequence. Quiescent galaxies are found 2\,dex or more below the main sequence, and the green valley region is situated between these two regimes. Our definitions are motivated by the apparent divisions between galaxy populations on color-magnitude diagrams of observed galaxies \citep[e.g.,][]{salim14}. 

Nearly all models, on the left frame (i.e., within $0.1R_{\rm vir}$) spend the majority of their evolution in the region we have defined here as the `green valley,' although visiting the quenched region multiple times around $z\sim 2-1$.  Only the $\epsilon_{\rm 50,LBH}$ model remains in the quenched region. All other AGN models, apart from the $\epsilon_{\rm 5,LBH}$ (an outlier),  recover and re-approach the main sequence by $z=0$. All models appear to evolve horizontally along the $x$-axis, i.e., roughly along the main sequence, with sSFR/sSFR$_{\rm MS}\sim 1$, until they reach $z\sim 2$. At this point the sSFR/sSFR$_{\rm MS}$ rapidly, and nearly vertically, decreases.  After $z\sim 1$, the sSFR/sSFR$_{\rm MS}$ begin to increase, again approaching the main sequence, with the final value on the $y$-axis scaling roughly inversely with the feedback strength, $\epsilon$.  The width of the U-shape corresponds to an extended period of time, $\sim 2$\,Gyr, without any major mergers, but the minor and intermediate mergers continue.  This trend exists {\it only} for the AGN galaxies, both EBH and LBH $\epsilon_0$ models do not experience this U-shape behavior --- their SFRs display a slow monotonic decrease in SFR after $z\sim 3-2$, as shown in Figure\,3 of Paper\,II. After analyzing the snapshots in Paper\,II, we concluded that the reason for this evolution of the jet models is a much stronger interaction between the jet and the perturbed gas in these models.  
 
Because it is difficult to estimate the galaxy size based on $0.1R_{\rm vir}$ observationally, we turn instead to the half mass radius, $R_{1/2}$, and half light radius, $R_{\rm e}$. The right-hand frame of Figure\,\ref{fig:sfms} displays the sSFR-$M_*$ relation, using the total mass inside $R_{1/2}$. The main-sequence relation from \citet{Popesso23} is based on several observational studies that use the total or aperture-corrected SFRs and stellar masses, rather than measurements within a fixed radius, such as $R_{1/2}$ or $R_{\rm e}$. Our measurements are not aperture-matched to the observational definition (as discussed in Section\,\ref{sec:num}) and may systematically under- or over-estimate the total sSFR by comparison. The difference in our chosen radius definition contributes to the offset toward lower sSFRs seen in the left-hand panel of Figure\,\ref{fig:sfms} in comparison to the right-hand panel. In this case, measurements within $R_{1/2}$ likely provide a closer proxy for the integrated quantities used in observational main-sequence studies such as in \citet{Popesso23}.

Comparing both frames of Figure\,\ref{fig:sfms}, we observe that, in the right frame, all models spend more time in the star-forming region prior to $z\sim 2$ relative to the same trajectories inside $0.1R_{\rm vir}$, indicating that SF is concentrated in the central $\sim 1-2$\,kpc at these early times. The final positions of our models on the right frame of this Figure still scale with $\epsilon$. However, rather than approaching the SF main sequence at $z=0$, the models are dispersed along the $y$-axis. While $\epsilon_0$ models end up at the SF main sequence, the AGN models end up substantially lower, with $\epsilon_{50}$ models finishing deep inside the quenched region, and even $\epsilon_{15}$ models sitting in this region. This behavior of sSFR within $R_{1/2}$ is in a sharp contrast with its behavior inside $0.1R_{\rm vir}$. Hence, importantly, the SF occurs predominantly outside $R_{1/2}$ after $z\sim 2$ in all AGN models (with the exception of $\epsilon_{\rm 5,LBH})$, and it is the jet feedback which is responsible for this effect. 

Next, we turn to Figure\,\ref{fig:Msigma} and address the evolution on the M$_{\bullet}-\sigma_{\rm bulge}$ plane, where $\sigma_{\rm bulge}$ is the 3-D stellar velocity dispersion of the stellar bulges, defined in Papers\,I and II.  The bulge parameters have been determined through kinematic decomposition carried out for every 30\,Myr of galaxy evolution. Following \citet{bi22a}, we have defined the bulge stellar population with $|j_{\rm z}/j_{\rm c}| < 0.5$, where $j_{\rm z}$ and  $j_{\rm c}$  are the specific angular momenta around the $z$-axis and the circular angular momenta respectively, within $0.1R_{\rm vir}$. For $\sigma_{\rm bulge}$ we have removed contribution from the rotational velocity, following \citet{stewart22}. 

Note that for both the EBH and LBH models, the $\sigma_{\rm bulge}$ values follow the same general trend of initial increase and subsequent decrease, approaching the median fit line from \citet{kormendy13} at low redshift. The 'knee' where the evolution changes from increasing to decreasing $\sigma_{\rm bulge}$ occurs around $z\sim 2$. All of the $\epsilon_{15}$ and $\epsilon_{50}$ models spend time within the 0.5\,dex error band from the median for a substantial amount of time, i.e., $z\ltorder 1$, while the values of both $\epsilon_5$ models nearly always lie outside this band. Interestingly, the AGN models converge to the $M_\bullet-\sigma$ relation, however their trajectories approach the relation nearly perpendicularly. 

Finally, the evolutionary tracks of LBH and EBH models differ profoundly on the $M_\bullet-\sigma$ plane, with EBH lying closer to the median at any given time --- a clear signature of the SMBH seeding time effect. Already at high $z$, both families are offset horizontally, with LBH galaxies having much larger bulge dispersion velocities, $\sim 180\,{\rm km\,s^{-1}}$, versus  $\sim 50\,{\rm km\,s^{-1}}$ for EBH.
This also reflects the subsequent evolution of stellar bulges --- they have slightly smaller masses and substantially larger radii for the EBH models.  

\subsection{Kennicutt-Schmidt scaling relation} 
\label{sec:KS}

\begin{figure*}[ht!]
\center
\includegraphics[width=0.8\linewidth]{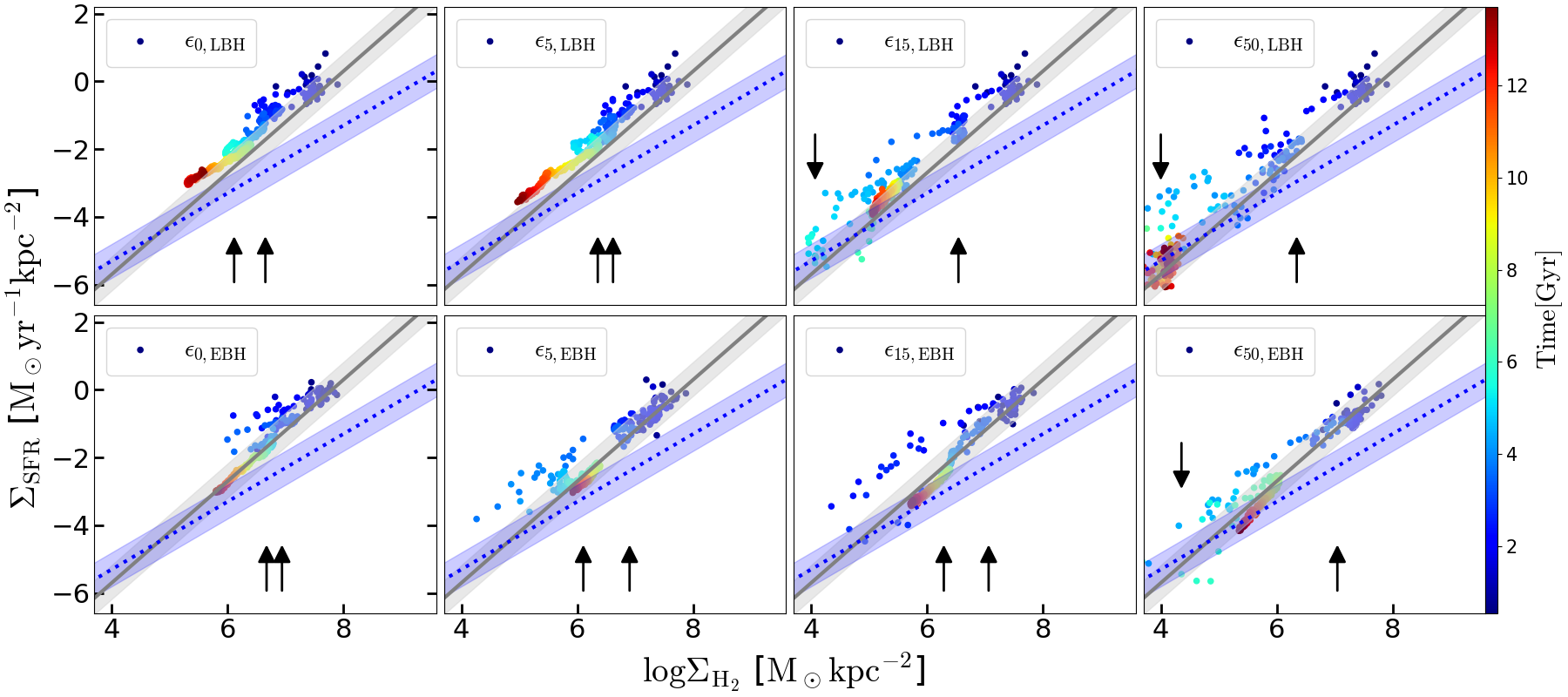}
\caption{$\Sigma_{\rm SFR}$ vs $\Sigma_{\rm H_2}$ for all gas inside $0.1R_{\rm vir}$. $\Sigma_{\rm SFR}=A(\Sigma_{\rm H_2})^{n}$ with depletion times of $\tau_{\rm dep} = 2$\,Gyr and slope $n=1$ \citep[e.g.,][]{bigiel08,leroy13} is given by dotted line, while $\tau_{\rm dep} = 0.5$\,Gyr and $n=1.5$ \citep[e.g.][]{genzel15,scoville17,tacconi18} is given by the solid line. The color of the markers indicate time. The shaded regions represent $\pm$0.5 dex around the relations. There is a single data point for each 30\,Myrs of evolution during $z=9-0$. The black arrows indicate $z=2$ and $z=1$.
\label{fig:KS}}
\end{figure*}

\begin{figure*}[ht!]
\center
\includegraphics[width=0.8\linewidth]{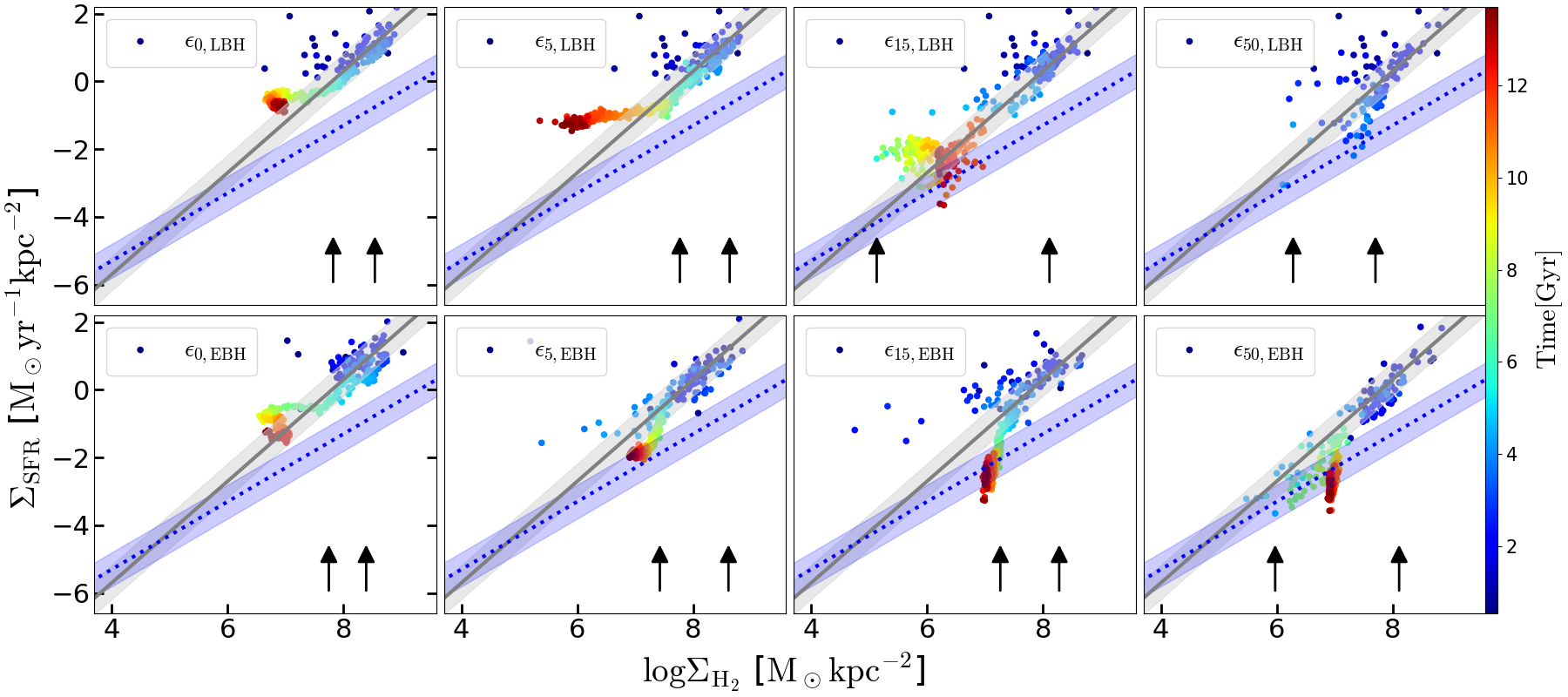}
\caption{Same as Figure\,\ref{fig:KS}, but using gas and stars within $R_{1/2}$ only. The black arrows indicate $z=2$ and $z=1$ (except for $\epsilon_{\rm 50,LBH}$ where the left arrow shows $z\sim 1.5$ and $\epsilon_{\rm 50,EBH}$ where the left arrow indicates $z\sim 1.3$ -- these models have no gas inside $R_{1/2}$ at $z=1$, so the arrows show the closest time with at least one gas particle in the region).
\label{fig:KS_Rhalf}}
\end{figure*}

\begin{figure}[ht!]
\center
\includegraphics[width=0.8\linewidth]{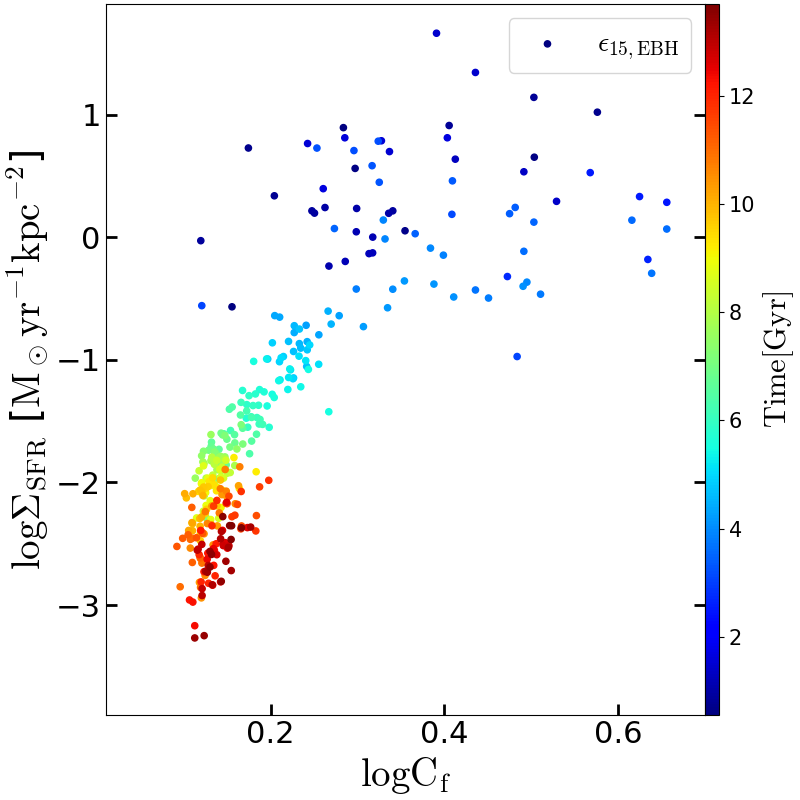}
\caption{Star formation rate surface density, $\Sigma_{\rm SFR}$, versus `clumping factor,' $C_{\rm f}$ (see definition in the text) within $R_{1/2}$ for the $\epsilon_{\rm 15,EBH}$ model. Marker color represents time in Gyr.
\label{fig:clump}}
\end{figure}

As a next step, we explore the evolution of our galaxies along the Kennicutt-Schmidt (K-S) relation.  Each of the models in Figure\,\ref{fig:KS} is shown in a separate frame, with data points spaced 30\,Myrs apart, during $z=9-0$. Marker color indicates time evolution with red showing later times and deep blue showing times just a few hundred Myr after the Big Bang. The black arrows represent $z\sim 2$ and $z\sim 1$, with time evolution moving from top right to bottom left. The solid gray and dashed blue straight lines show fits determined by observations (discussed further below) of the form $\Sigma_{\rm SFR}=A{(\Sigma_{H_2})^n}$, where $A=1/\tau_{\rm dep}$ represents the gas depletion rate.  

We adopt two different values for each of the parameters for these fits representing differing SF scenarios.  The dashed line uses a depletion time of 2\,Gyr and $n=1$. Values in this range have been found in empirical fits to the molecular K-S law from a wide array of observational and numerical studies of local, 'normal' disk galaxies \citep[e.g.,][]{bigiel08,leroy13,semenov19}.  The solid line uses a depletion time of 0.5\,Gyr and $n=1.5$. This steeper slope and shorter depletion time is generally indicative of rapid SF, as at high-$z$ or in starburst galaxies \citep[e.g.][]{genzel15,scoville17,tacconi18}. 

The galaxy evolutionary tracks on Figure\,\ref{fig:KS} follow the K-S relation with $\tau_{\rm dep} = 0.5$\,Gyr and $n=1.5$ (the solid line), while clearly disagreeing with the alternative $\tau_{\rm dep} = 2$\,Gyr and slope $n=1$ dashed line. During $z = 9-0$ time period, the modeled galaxies also move to lower $\Sigma_{\rm H_2}$ and consequently to lower $\Sigma_{\rm SFR}$. This indicates that the galaxies reduce in surface density of molecular gas and slow down in SFR with time. Due to the convergence between the solid and dashed lines at lower $\Sigma_{\rm H_2}$, the models, and especially the AGN models, satisfy both  $\Sigma_{\rm SFR} - \Sigma_{\rm H_2}$ K-S laws at low redshifts. Notably, the AGN models with larger jet feedback lie lower on the $y$-axis at lower redshifts --- this is especially prominent in the LBH models. This indicates that the jet feedback slows down SFR.

\begin{figure*}[ht!]
\center
\includegraphics[width=0.8\linewidth]{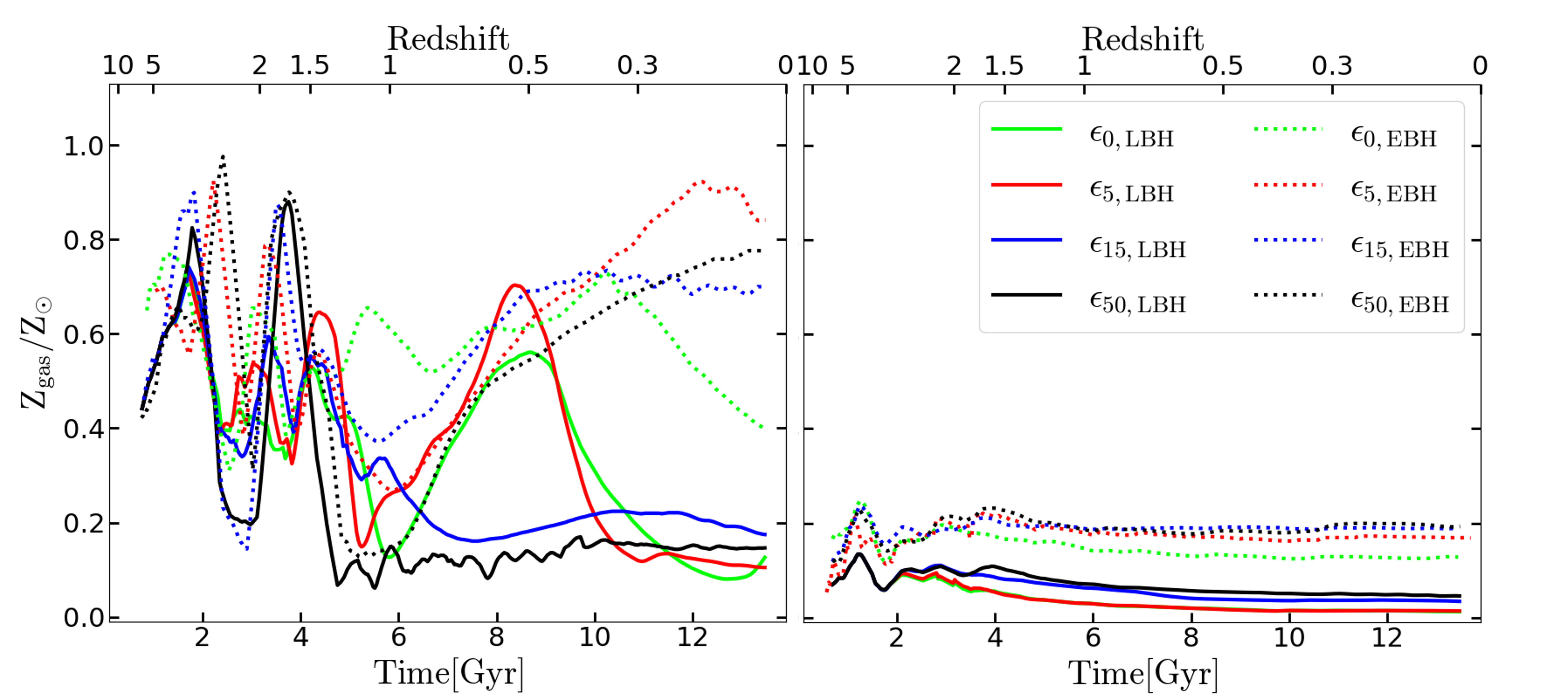}
\caption{Evolution of the mass-averaged gas metallicity in modeled galaxies. \textbf{\textit{Left:}} the ISM, represented by all gas inside $0.1R_{\rm vir}$. \textbf{\textit{Right:}} the CGM, represented by all the gas between 0.1$R_{\rm vir}$ and $2R_{\rm vir}$.
\label{fig:aveZ}}
\end{figure*}

In both $\epsilon_{50}$ models and in $\epsilon_{\rm 15,LBH}$ model, the $z=1$ arrow is found in the bottom-left corner, as there is very little gas, and thus the SFR is low at this time. The gas is evacuated due to jet-ISM interactions and subsequent expanding shocks which push out the gas (see paper II). 

Figure\,\ref{fig:KS_Rhalf} exhibits the same K-S relationship as Figure\,\ref{fig:KS}, but for the half-mass radius, $R_{1/2}$. Again, the black arrows show $z\sim 2$ and $z\sim 1$. However, for both $\epsilon_{50}$ models, the gas is absent inside $R_{1/2}$ at $z=1$. So, instead of $z=1$, the second arrows are shown at $z\sim 1.5$ for $\epsilon_{\rm 50,LBH}$, and $z\sim 1.3$ for $\epsilon_{\rm 50,EBH}$.

Using $R_{1/2}$ as a size benchmark, we observe a number of differences with Figure\,\ref{fig:KS}. First, the evolutionary tracks tend to terminate at higher $\Sigma_{\rm SFR}$ at low redshifts, meaning that $\Sigma_{\rm SFR}$ in the central regions is higher, compared to $0.1R_{\rm vir}$, especially in the $\epsilon_{0}$ and $\epsilon_{\rm 5,LBH}$ models. So, while sSFR is lower in the central 1-2 kpc of the AGN-models (as discussed in Section \ref{sec:figure2}), the $\Sigma_{\rm SFR}$ is higher in this region. Second, evolutionary trajectories behave differently in the LBH and EBH models at low redshifts. Both horizontal and vertical evolutionary features are observed. Especially prominent among these are the horizontal evolution during $z\sim1-0$ in $\epsilon_{\rm 5, LBH}$ and the vertical `fingers' after $z\sim1$ in  $\epsilon_{\rm 15,EBH}$ and $\epsilon_{\rm 50,EBH}$.

The simplest reason for a sharp decline in SFR at fixed  $\Sigma_{\rm H_2}$ can be related to our recipe for SF, namely, that stars form only where and when the gas is self-gravitating, i.e., when the virial parameter, $\alpha_{\rm vir} < 1$. The virial parameter depends on the gas sound speed, velocity gradient, and density, and hence the SF is dominant in $H_2$. The $H_2$ content is estimated based on the total gas surface density and metallicity within the kernel \citep{krumholz11}, and thus selects gas that meets the density and self-gravity requirements for the SF, leading to the observed correspondence with the K-S relation. Surprisingly, as discussed above, with the jet feedback, we find decreasing SFR without a substantial change to the $H_2$ content.


To understand this behavior, we have measured the degree of clumpiness in the cold gas, $T<10^4$\,K, inside $R_{1/2}$. We define the clumping factor as
\begin{equation}
    C_{\rm f}=\sqrt{<\rho^2>/<\rho>^2},
\end{equation}
where average gas density is $ <\rho>=\Sigma_{\rm i}( \rho_{\rm i} V_{\rm i}) / \Sigma_{\rm i} V_{\rm i},$
the volume is $V_{\rm i} = 4/3 \pi h_{\rm i}^3$, and $ <\rho^2>=\Sigma_{\rm i}( \rho_i^2 V_{\rm i}) / \Sigma_{\rm i} V_{\rm i}.$

Here $\rho_{\rm i}$ is the kernel-weighted gas density of each element, $h_{\rm i}$ is the kernel radius, and the brackets represent volume-weighted averages. The clumping factor expresses the degree of non-uniformity. $C_{\rm f}$=1 corresponds to a perfectly homogeneous gas distribution, while $C_{\rm f}>1$ indicates the buildup of dense substructures or clumps within the ISM. 

An example of the clumping factor, $C_{\rm f}$, dependence on $\Sigma_{\rm SFR}$ within $R_{1/2}$, model $\epsilon_{\rm 15,EBH}$ is shown in Figure\,\ref{fig:clump}. We confirm that in all models, but especially in the AGN models with a stronger jet feedback, a sharp decline in the $C_{\rm f}$ coincides with a sharp decline in the SFR. The appearance of the vertical `finger' in Figure\,\ref{fig:KS_Rhalf} corresponds to reduced SFR in the cold gas when the gas becomes more uniform. 

Furthermore, four models in Figure\,\ref{fig:KS_Rhalf}, i.e., both $\epsilon_0$ models, $\epsilon_{\rm 5,LBH}$, and $\epsilon_{\rm 15,LBH}$, display horizontal `fingers' at low redshifts, i.e., $\Sigma_{\rm SFR}$ remains the same despite a decrease in $\Sigma_{\rm H_2}$ by about an order of magnitude. These features coincide with the appearance of transient stellar bars in the central few kpc of these models, and we find one-to-one correlation between appearance of these fingers and bars. Note that bars or oval distortions appear when the jet power remains below $L_{\rm jet}\sim 10^{41}\,{\rm erg\,s^{-1}}$ after $z\sim 1.5$, with the exception of $\epsilon_{\rm 50,LBH}$ which lacks gas in the center of the galaxy during this time. As typical of such strong bars, they push the gas out across the co-rotation, and funnel the gas within their radii towards the center, where SF is spiked, both in simulated bars and in observed galaxies \citep[e.g.,][]{devereux87,athana92,knapen95,knapen00,jogee02,bi22b}.

\subsection{Gas phase mass-metallicity relation} 
\label{sec:MZR}

Next, we turn to the evolution of gas-phase metallicity within the modeled galaxies, which together with stellar mass of galaxies represent the most fundamental signatures of their evolution. Before discussing the evolution of our galaxies on the Mass-Metallicity relation, we compare the evolution of the ISM and CGM metallicities. Figure\,\ref{fig:aveZ} exhibits the evolution of the average gas metallicity, $Z_{\rm gas}$, in the ISM (left frame) and CGM (right frame). In both the ISM and CGM frames, the separation between EBH and LBH metallicity curves becomes significant around $z\sim 5-4$. Hence, it appears concurrent to the major merger happening at this time. Thereafter, the separation between the EBH and LBH CGM curves increases until $z\sim 1.5$. After this, the EBH metallicity remains flat, while in the LBH models we see a slow decay until $z\sim 0.5$, where it becomes flat. 

In the CGM, the cause of such an early difference in metallicities between the EBH and LBH models can result from the SN feedback, being stronger in the former models, and the early presence of the jet feedback in these models due to the seeding time of the SMBH --- both can be amplified during a major merger. A plausible scenario is that some of the early enriched gas is pushed into CGM by this feedback in the EBH models, and then falls back, as discussed in Paper\,II.

In the ISM, the evolution is more complicated because it is subject to competing processes --- the SF, expulsion of the enriched gas into the CGM, and influx of both pristine and enriched gas from the CGM. Fluctuations up to 1\,dex in the average metallicity occur on timescales of $\sim 1$\,Gyr in both LBH and EBH models, which is characteristic of the DM halo crossing time; after $z\sim 1$ these fluctuations slow down. At $z\ltorder 0.5$, a substantial difference between the EBH and LBH models appears, with the former remain enriched by about 0.2-0.8\,dex compared to the LBH models. Comparing metallicities of the ISM and CGM at low redshifts, the ISM is more metal-rich by a factor of $\sim 3-4$.

\begin{figure*}[ht!]
\center
\includegraphics[width=0.8\linewidth]{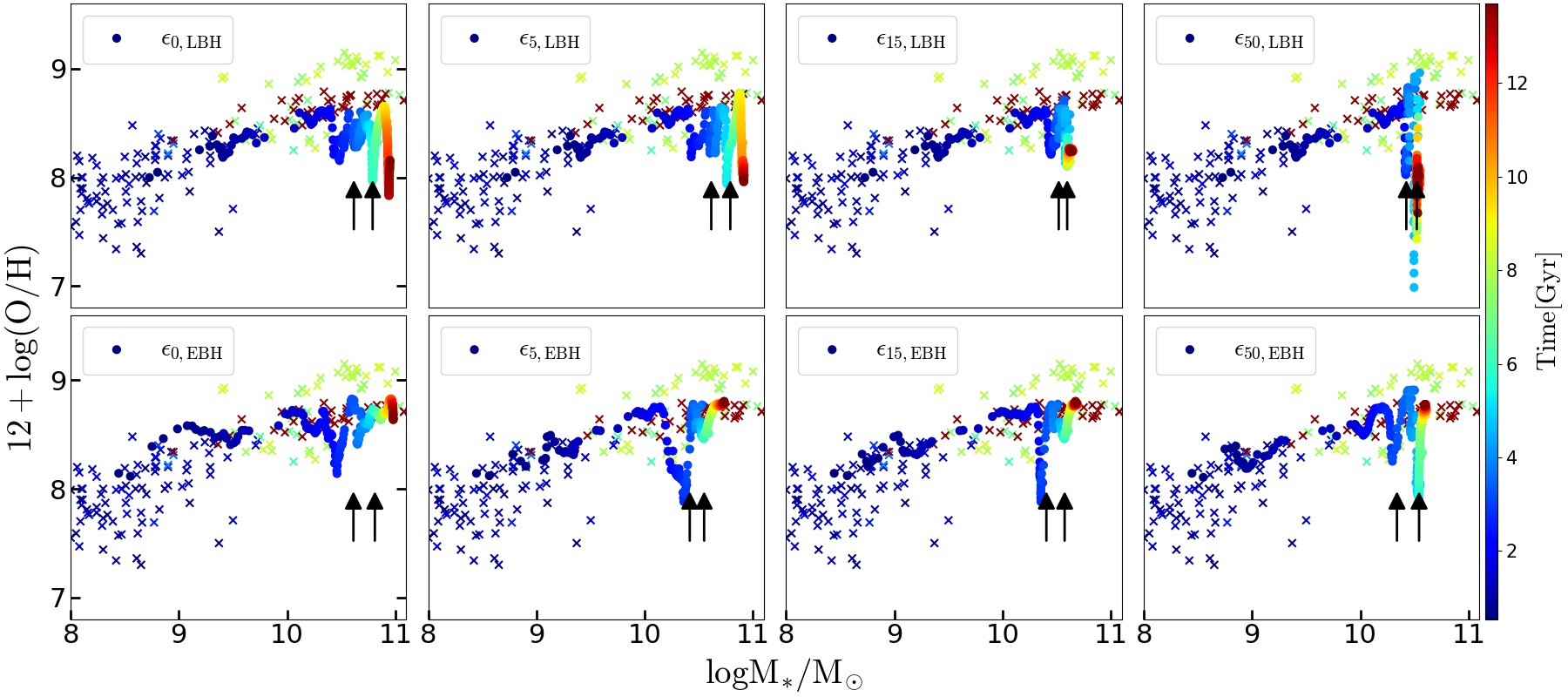}
\caption{Evolution of gas phase metallicity as 12+log(O/H) vs stellar mass for gas and stars inside $0.1R_{\rm vir}$ for modeled galaxies.  The color of the markers indicate time with a single data point for each 30\,Myr of evolution during $z=9-0$. The dots represent the modeled galaxies, and crosses represent the observational data for individual galaxies from \citet{sanchez2013, maier2015,lewis2024,sarkar25,scholte25}. The black arrows show $z=2$ and $z=1$.  The large wiggles/vertical fingers in our models represent major influx/outflow of lower/higher metallity gas from the CGM.
\label{fig:MZR}}
\end{figure*}

Figure\,\ref{fig:MZR} displays the mass-metallicity relation (MZR) versus stellar mass, $M_*$. For each 30\,Myr we have computed the gas-phase oxygen abundance, 12+log(O/H), by taking a mass-weighted average of O and H across the gas elements {\it within the chosen galaxy aperture,} i.e., $0.1R_{\rm vir}$ or $R_{1/2}$. This $M_*-Z_{\rm gas}$ relation for modeled galaxies has been compared to observed galaxies, where a variety of techniques have been employed \citep[e.g.,][]{sanchez2013,maier2015,lewis2024,sarkar25,scholte25}. The most direct approach, the $T_{\rm e}$ method, uses faint auroral lines, such as [O III] $\lambda$4363, to derive the electron temperature, from which ionic abundances are computed \citep[e.g.,][]{curti17}. However, these lines are weak in metal-rich systems and at high $z$, where bright-line ratios are used instead.

In Figure\,\ref{fig:MZR}, for $z\gtorder 2$, our models follow the observations. For lower $z$, the evolution of model galaxies flattens (but see the discussion related to Figure\,\ref{fig:MZR_Rhalf} below) and exhibits 1--3 vertical `fingers,' meaning that metallicity changes up to 2\,dex for nearly fixed $M_*$. These sudden changes in the metallicity can be directly related to low-metallicity gas influx and to significant drops in the galaxy gas content, which is evacuated due to jet-ISM interactions. Subsequent increases in the metallicity follow from SF enriching the accreted material and to the influx of a higher metallicity gas previously expelled into the CGM. These 'fingers' are narrower at later times, when $M_*$ growth slows considerably --- thus changes appear abrupt on the mass scale.  These fluctuations in average galaxy metallicity in LBH and EBH galaxies are visible also in Figure\,\ref{fig:aveZ} (left frame), where the timescale is better represented. In short, the presence of jet feedback influences the duration and frequency of these disruption events, as we elaborate in section\,\ref{sec:discussion}.

Figure\,\ref{fig:MZR_Rhalf} shows the 12+log(O/H) metallicity versus stellar mass inside $R_{1/2}$. Metallicity of modeled galaxies in this Figure are higher at $z\ltorder 5$. For $z > 5$, the modeled metallicity follows the metallicity in {\it observed} galaxies. The vertical `fingers' do not extend to metallicities as low as in Figure\,\ref{fig:MZR}. This difference between Figures\,\ref{fig:MZR} and \ref{fig:MZR_Rhalf} confirms that there is a negative gradient in the gas metallicity across the modeled galaxies, with the central regions being more enriched at all redshifts, except where the cold, metal-rich gas fraction is greatly reduced or absent in the central regions, resulting in the downward 'fingers' on the diagram. 

After $z\sim 0.5$, a decrease in 12+log(O/H) can be observed for our LBH models relatively to the EBH models. At least in part, it is due to the differences in the SN feedback, as we see distinction of $\sim 0.5$\,dex even between the $\epsilon_{\rm 0}$ models.  It is especially pronounced near $z=0$, where the EBH models tend to be $\sim 0.5$\,dex richer than the LBH models. In the LBH models, there are strong downward trending 'fingers,' during this time, as evident in Figure\,\ref{fig:MZR}, where accretion of low-metallicity gas is better represented by the larger aperture.  In Figure\,\ref{fig:aveZ}, we show that the EBH models have a more enriched CGM when compared with the LBH models, due in part to the earlier seeded SMBH, and amplified by the enhanced SN feedback. Evidently, this enrichment in the CGM is complimented by a similar enrichment of the ISM after $z\sim 0.5$.

\begin{figure*}[ht!]
\center
\includegraphics[width=0.8\linewidth]{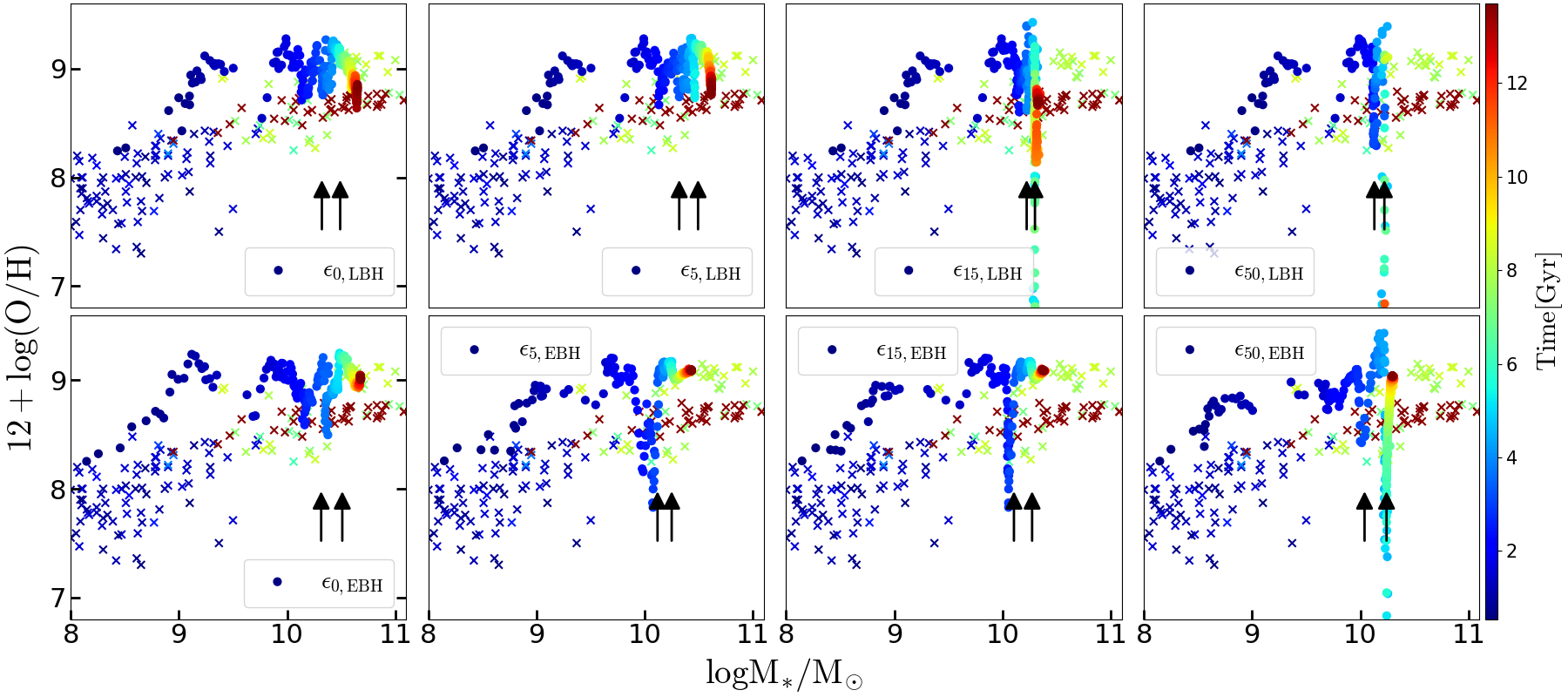}
\caption{Same as Figure\,\ref{fig:MZR}, but using gas and stars within $R_{1/2}$ only. Note the presence of metallicity gradients as the central regions of galaxies have higher average metallicities than those measured in $0.1R_{\rm vir}$ in Figure\,\ref{fig:MZR}.
\label{fig:MZR_Rhalf}}
\end{figure*}

\subsection{Baryonic Tully-Fisher Relation} 
\label{sec:BTFR}

\begin{figure}[ht!]
 \center
\includegraphics[width=0.8\linewidth]{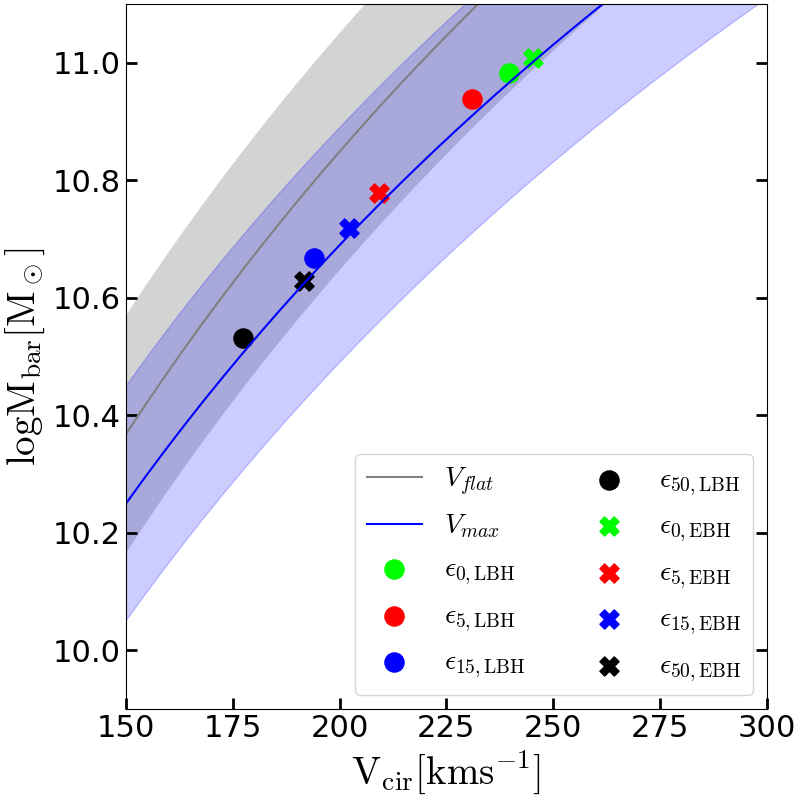}
\caption{Baryonic mass, $M_{\rm bar}$, inside $0.1R_{\rm vir}$ versus circular velocity, $V_{\rm cir}$, at $0.1R_{\rm vir}$ at $z=0$ for modeled galaxies.  The solid lines show empirical fits from \citet{lelli19} using different velocity definitions with shaded regions representing $\pm 0.2$\,dex around these lines.
\label{fig:btfr}}
\end{figure}

We now turn to the baryonic Tully-Fisher Relation (BTFR) for modeled galaxies at $z=0$. We adopt the circular velocity measured at the galaxy edge, $0.1R_{\rm vir}$, which typically encloses the stellar and cold gas components of the galaxy and probes the regime where baryons begin to dominate the gravitational potential. This provides a stable and physically motivated velocity scale that reflects the galaxy’s inner mass distribution without being overly sensitive to small-scale fluctuations or feedback-driven outflows, and works as a reasonable analogue to $V_{\rm flat}$ measured in observational BTFR studies. We compare these to the empirical BTFR derived using the $V_{\rm flat}$ and $V_{\rm max}$ (i.e., the peak of the rotational velocity), velocity definitions for local galaxies \citep[e.g.,][]{lelli19} in Figure \ref{fig:btfr}. The shaded regions show $\pm 0.2$\,dex and represent the approximate scatter around the relation from observed galaxies. In each case, we track the baryonic mass within $0.1R_{\rm vir}$. 

There are systematic differences in the position of  modeled galaxies on the BTFR as a function of jet feedback. The lowest feedback strengths, i.e., $\epsilon_0$ and $\epsilon_{\rm 5,LBH}$, steadily increase in both baryonic mass and $V_{\rm  cir}$ at $0.1R_{\rm vir}$, converging at late times to the observed relation and approaching $V_{\rm cir}\sim 225–250\,{\rm km\,s^{-1}}$. Higher feedback strengths, i.e., $\epsilon_{\rm 5,EBH}$, and both $\epsilon_{15}$ and $\epsilon_{50}$, gain baryons more slowly and achieve a systematically lower final circular velocity of $V_{\rm cir}\sim 180-210\,{\rm km\,s^{-1}}$. This trend indicates that stronger jet feedback reduces central baryonic concentrations, 
flattening the mass–velocity growth (Papers\,I and II).

\begin{figure}[ht!]
\center
\includegraphics[width=1.0\linewidth]{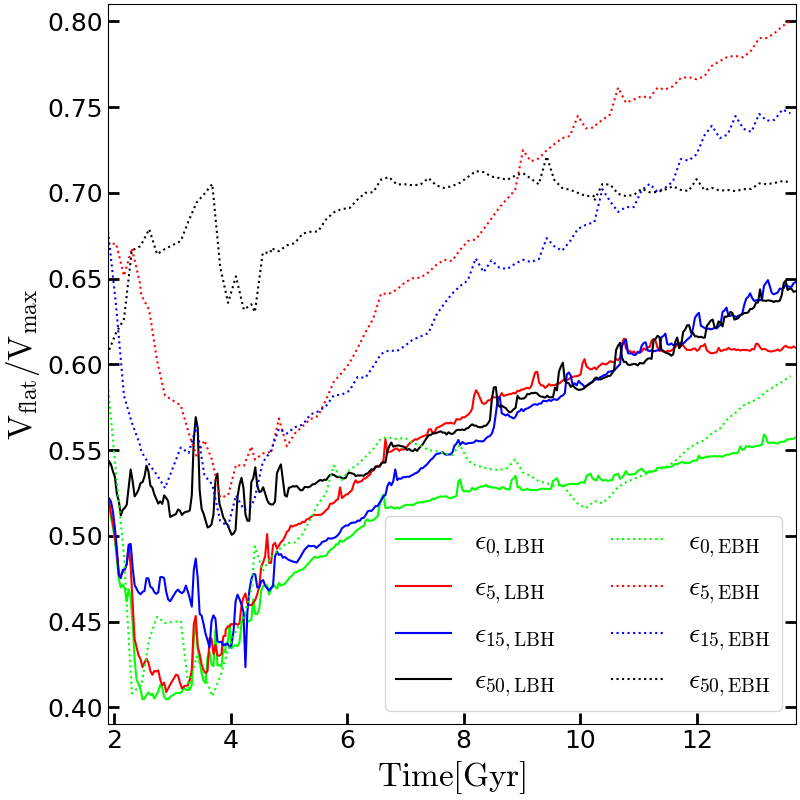}
\caption{Evolution of $V_{\rm flat}/V_{\rm max}$ for modeled galaxies. $V_{\rm flat}$ is the circular velocity measured at $0.1R_{\rm vir}$ and $V_{\rm max}$ is the maximum value of $V_{\rm cir}$ between the galaxy center and $0.1R_{\rm vir}$. This ratio represents the compactness of the matter distribution in the central regions of the halo.
\label{fig:conc}}
\end{figure}

Figure\,\ref{fig:conc} displays $V_{\rm flat}/V_{\rm max}$ versus time, where $V_{\rm flat}$ is the circular velocity measured at $0.1R_{\rm vir}$ and $V_{\rm max}$ is the maximum value of $V_{\rm cir}$ between the galaxy center and $0.1R_{\rm vir}$. The ratio $V_{\rm flat}/V_{\rm max}$ provides a useful proxy for the galaxy central concentration and compactness. We find that models including SMBH jet feedback consistently show higher values of this ratio, indicating less centrally concentrated mass distributions compared to the non-AGN cases. This is consistent with the Seyfert feedback acting to heat, expel, or prevent the inflow of low angular momentum gas, thereby suppressing the buildup of dense central baryonic components. 

For both the LBH and EBH models the evolution of the $\epsilon_{50}$ model is flatter, indicating that strong jet activity limits central mass assembly throughout cosmic time rather than only at late epochs. Despite this difference, all Seyfert models of similar seeding time converge toward comparable concentration values by $z=0$, with the EBH Seyfert models ending up $\sim 15$\% less compact than the LBH Seyferts. This suggests that the late-time growth of the outer disk and DM halo potential dominates over earlier feedback-driven differences, while an earlier SMBH seeding time leads to a substantial and sustained difference in the galaxy compactness.

\section{Discussion and conclusions} 
\label{sec:discussion}

We have explored how Seyfert jet feedback influences the evolutionary tracks of modeled galaxies on a number of major scaling relations during $z\sim 9-0$. A suite of galaxies has been modeled using high-resolution zoom-in cosmological simulations (Paper\,I and II), and the model parameters have been varied as described in section\,\ref{sec:num}, and summarized in Table\,\ref{tab:galprops}. Seyfert models are compared to control models with no SMBH, to isolate the effects of the jet feedback.

Our results show that jet feedback in Seyfert galaxies can introduce systematic and sustained differences in the galaxy's position on major scaling relations. In this section we analyze the causes and implications of our results within the greater context of observed galaxies and numerical simulations. We also discuss how our numerical results can be linked to the large-scale observable photometric, spectroscopic, and structural signatures detectable by multi-wavelength observations of Seyfert galaxies.

\subsection{Comparison with Observations, Simulations, and Theory} 
\label{sec:compare}

\underline{\bf M$_{\rm halo}-$M$_{*}$ Relation.} Our results show that Seyfert jet feedback plays an important role in suppressing host galaxy stellar mass growth. Modeled jetted galaxies are consistently shifted downward on the M$_{\rm halo}-$M$_{*}$ plane relative to the no-SMBH models, aligning more closely with the median (Fig.\,\ref{fig:smbh}). This effect is amplified in simulations with higher jet power, where feedback more effectively slows down the SF. Without SMBH feedback, the galaxy sits near the upper limit of the relation, producing up to $3\times M_*$ for its halo mass by $z=0$.   

While AGN feedback is conjectured to be a major contributor to SF quenching in massive galaxies and clusters, we show that mechanical and thermal jet feedback is also effective in reducing SF in less massive Seyfert-type galaxies. Sufficient evidence exists that all massive galaxies host SMBHs, and that 'regular' or ‘star-forming’ galaxies are merely hosts to dormant or obscured SMBHs \citep[e.g.,][]{harrison24}. Our work shows that the presence of Seyfert activity can be necessary in this intermediate-mass regime to regulate the SF.

Jet feedback affects the ISM and CGM by depositing thermal energy and linear momentum in the gas. We showed in Papers I and II that less powerful jets are trapped within the central few kpc, and generate hot expanding bubbles there, while more powerful jets reach the CGM, inducing turbulence and affecting the SF. Expanding shocks drive enriched gas from the galaxy, which cools down and leads to SF in the inner CGM. The expanding gas flow displays a roughly bi-conical symmetry, depending on the angle of the jet with respect to the galactic disk. One would expect cooling shells with SF, which in principle could be detected using JWST spectroscopy. To the extent that the expelled ISM will mix with the CGM gas, one would expect pockets of enriched CGM gas. Therefore, jets have a dual effect on the SF, by reducing it in the ISM and inducing it in the inner CGM. Overall, while the Behroozi relation shows only a weak dependence with redshift, the modeled galaxies experience a stronger evolution, due partly to merger events and interactions. They depart from the median until $z\sim 2$, i.e., the halo assembly time, and this departure scales inversely with the jet feedback. Galaxies without SMBH show the largest departure from the median, which decreases sharply towards $z\sim 1$, and stays about constant thereafter.     

\underline{\bf sSFR$-$M$_*$ Relation.} In our simulations, jet feedback reduces not only the SFRs but also the sSFR, causing an occasional shift towards the red sequence. This suppression is most pronounced during and after the transitional phase, $z\sim 1-2$, where Seyfert models dip substantially below the star-forming main-sequence (SFMS). This rapid reduction comes from jet-ISM interactions that drive expanding shocks, leading to reduction of a cold gas in the galaxy --- we observe a strong correlation between the SFR and the amount of H$_2$ in galaxies \citep[][for details]{goddard25b}.  Interestingly, for $z\ltorder 1$, all Seyferts begin to converge towards the observed SFMS (see Figure\,\ref{fig:sfms}), though their final positions scale inversely with the feedback strength. This trend suggests that jet feedback regulates sSFR without completely quenching SF in jetted Seyferts.  

Most observed Seyfert jets are hosted by actively star-forming galaxies \citep[e.g.,][]{Foschini2020, Varglund2022}. In principle, the presence of jets can have opposite effects on SFR \cite[e.g.,][]{silk24,jin25}, as has been confirmed by simulations and observation \citep[e.g.,][]{mukherjee18,mercedes-feliz23,caccianiga15,rao23,garcia-burillo24,Kurian2024}. 

Comparing the sSFR within $0.1R_{\rm vir}$ and $R_{1/2}$, the sSFR is reduced more inside $R_{1/2}$ during late-stage evolution. This is in agreement with observations that jets displace ISM in the central few kpc \citep[e.g.,][]{morganti15,Murthy2022,nandi23}, leading to lower central stellar surface densities \citep{lammers23,acharya24}. Simulations involving various types of feedback \citep{byrne24} and limited to jet feedback \citep{goddard25b} support this trend. 

\underline{\bf K-S Relation.} It has been proposed that AGN galaxies possess a steeper slope and hence shorter depletion time for the molecular K-S relation in comparison with `regular' star-forming galaxies due to a shock-triggered SF or to lower SF for fixed H$_2$ content because of induced turbulence \citep[e.g.,][]{salvestrini20,salome23}.  We confirm in Figure\,\ref{fig:KS_Rhalf} that the sudden decline in SFR expressed by vertical 'fingers' at low $z$ can indeed be interpreted by a shorter depletion time within $R_{1/2}$ in Seyfert models. We tested the possibility that the sharp decrease in the SFR and increase of the K-S slope can be associated with jet feedback reducing the clumpiness in H$_2$, as is evident in Figure\,\ref{fig:clump}.

\underline{\bf M$_\bullet-\sigma_{\rm bulge}$ Relation.} We observe that convergence of modeled jetted Seyferts to the median $M_{\bullet}-\sigma_{\rm bulge}$ depends on jet feedback and on the seeding time of the SMBH (Fig.\,\ref{fig:Msigma}). For LBH models, the weak feedback, $\epsilon_{\rm 5,LBH}$, is characterized by growing $\sigma_{\rm bulge}$ together with $M_\bullet$. Stronger feedback displays preferential growth of the SMBH only. For the EBH models, the trend is the same, but the tracks lie much closer to the median, meaning that $\sigma_{\rm bulge}$ is smaller by a factor of 2 at $z>2$. Because the masses of bulges at corresponding times of evolution are similar, it is larger bulge sizes in the EBH models that leads to smaller $\sigma_{\rm bulge}$ (Paper\,II). This agrees with our overall conclusion that earlier SMBH seeding time and stronger SN feedback result in puffed up stellar and gaseous components, both in the disk and bulge.

All evolutionary tracks in Figure\,\ref{fig:Msigma} diverge from the median before $z\sim 2$, after this they turn around and approach the median at steep angles. This behavior may be a signature that the $z=0$ values of $M_\bullet-\sigma_{\rm bulge}$ are transient and not convergent ones. \citet{taylor16} observed that galaxies can grow faster in $\sigma_{\rm bulge}$ than $M_{\bullet}$ during merger events, when $M_*$ grows rapidly, before the gas reaches the center to fuel the SMBH growth.  Similarly, \citet{kormendy13} suggested that galaxies converging from the left to the scaling relation do so as a result of mergers that convert disk mass to bulge mass in the absence of significant SMBH growth. Our models in Figure\,\ref{fig:Msigma} {\it always} converge to the median from the higher $\sigma_{\rm bulge}$, i.e., from the right side. Tracks that cross the median and reside on the left, imply a transient nature to this `convergence.' Thus, we observe significant epochs in evolution, when SMBHs and bulges do not lie on the $M_\bullet-\sigma_{\rm bulge}$ relation, and expect this outcome from observations as well.  
  
\underline{\bf M$_*-$Z$_{\rm gas}$ Relation.} In our simulations, metal enrichment arises solely from SN\,II, these deposit thermal energy, momentum, and metals into neighboring gas elements. Expanding shocks also advect the metals they entrain; their primary action is to stir, lift, and redistribute gas. While our feedback generates turbulence that mixes metals over resolved scales, sub-grid diffusion is absent, meaning that mixing may be patchier than in real ISM and CGM environments.

Observed metallicity values typically represent only ionized, star-forming regions, whereas the simulated metallicities represent a mass-weighted average over all cold and warm gas, including phases that are invisible to optical line diagnostics.  This can lead to systematic differences in simulation-based and observationally-inferred abundances, although they have been claimed to be statistically insignificant \citep[e.g.,][]{ma16}. 

At both low and high $z$, ratios of bright emission lines, like R$_3$ = [O III]/H$\beta$, 
serve as empirical or model-based calibrations for metallicity \citep[e.g.,][]{sanchez2013,maier2015,lewis2024,sarkar25}. \citet{scholte24} used JWST observations spanning $1.6\leq z\leq 7.9$, to measure faint auroral lines, revealing that strong-line calibrations are sensitive to the ionization parameter and abundance ratios, biasing metallicity estimates. These works show that the choice of calibration can shift metallicity estimates by a factor of 2–3, and is a major source of systematic uncertainty, resulting in large scatter in the MZR when data from multiple independent studies (see Fig.\,\ref{fig:MZR}). Here we comment on the influence of the jet power on the evolution of our galaxies on the MZR, and compare with observations that have been included here for reference.

A major challenge in studying the MZR for AGN host galaxies in observations is that AGN emission contaminates the diagnostic nebular lines used to infer gas-phase metallicities. The hard ionizing spectrum of an AGN introduces broad-line components and shock-excited emission. As a result, the standard strong-line ratios no longer trace HII region physics and instead reflect a mixture of AGN photo-ionization, shocks, and SF-driven emission. Because most metallicity calibrations assume pure HII region excitation, including AGN-contaminated spectra can bias inferred abundances, typically driving line ratios toward artificially high metallicities or producing ambiguous upper/lower–branch solutions. For these reasons, the majority of large MZR surveys exclude AGN.

Despite these difficulties, several targeted studies have derived metallicities for AGN hosts by attempting to account for AGN and star-forming emission separately. \citet{thomas19} found that local Seyfert galaxies demonstrate a 0.9\,dex upwards offset from the MZR measured for local star-forming galaxies, indicating that they are on average more metal rich. However, they also find that at least 0.3\,dex of this difference comes from the differing methods of determining 12+log(O/H) for the AGN versus non-AGN samples. \citet{li24} found a similar, but more modest increase in average oxygen-abundances for AGN hosting-galaxies with log\,$M_*/M_\odot<10^{10.5}$ and show that this change also correlates with lower average SFRs. This echoes our result that at late times the stronger feeback, EBH models tend to be more metal rich and SFRs are reduced in our Seyferts vs non-Seyferts galaxies.

\citet{nascimento2022} demonstrate that while nearby AGN-hosting galaxies from a MaNGA sample typically have negative metallicity gradients. \citet{amiri24} observe the same phenomenon in NGC\,7130. On the other hand, \citet{armah24} find that 87\% of their Seyfert galaxy sample display a positive metallicity gradient. While we do not observe positive metallicity gradients in our galaxies, our Seyferts display a flatter gradient at $z=0$ \citep{goddard25a}. 

\underline{\bf Extragalactic Jet-triggered Shocks.}
We find that all AGN models display jet-triggered shocks that propagate to Mpc scales with initial velocities $\sim 3,000\,{\rm km\,s^{-1}}$ decelerating to $\sim 200\,{\rm km\,s^{-1}}$, and result in over-pressured 'bubbles' in the CGM and IGM (Papers\,I and II). Faster shocks correspond to stronger jets, high $z$ and initial stages of bubbles. While these are not directly measured by any of the major galaxy scaling relations, they are directly responsible for many of the evolutionary changes in the galaxy that {\it are} measured by the scaling relations and thus cannot be neglected in this discussion. Shocks are discussed in detail in Paper\,II, and their timescale and observable signatures are discussed further in Sections\,\ref{sec:time} and \ref{sec:observation}. Here we present only context for these shocks from recent literature.

Shocks are quantified via spatially resolved H$_2$ and ionized gas kinematics, line ratios, and energetics that tie the outflow to the radio jet rather than to a disk wind \citep[e.g.,][]{Tadhunter2014,morganti15}.  Previous simulations have studied the generation of shocks due to jet-ISM interactions by directly modeling jets in isolated disk galaxies \cite[e.g.,][]{wagner11,wagner12}. However, previous observational and simulation studies of jet–ISM interactions in Seyferts have focused on galaxy scales, where jets drive fast, multiphase outflows and produce shock signatures seen in local Seyferts such as IC\,5063 \citep{Tadhunter2014} and NGC\,4258 \citep{cecil00}. These works have demonstrated how low-power jets couple efficiently to a clumpy ISM, accelerating molecular, ionized, and X-ray emitting gas and generating strong line-ratio and kinematic signatures, but they do not follow the shock front or hot bubble beyond the galactic disk. Our results show the long-term effects of these shocks, and how they leave imprints on global galaxy scaling relations that cannot be measured from ISM-scale studies alone. In doing so, our models provide a self-consistent view of how Seyfert jets influence both the galaxy and its baryon cycle.

\underline{\bf BTF Relation.} The key physical driver of changes in the evolution of the BTFR in our simulations is the jet feedback which regulates central baryonic mass assembly. In the no-SMBH and $\epsilon_{\rm 5,LBH}$ models, gas cooling remains efficient, and the galaxy continues to accumulate mass inside $0.1R_{\rm vir}$. This deepens the central potential and increases $V_{\rm cir}$. In contrast, stronger jets heat and expel gas from the inner halo, suppressing the supply of cold star-forming material and preventing further deepening of the potential well. As a result, the galaxy accumulates baryonic mass more slowly and reaches a lower final $V_{\rm cir}$.  

All galaxies, independent of feedback strength, become more concentrated from early times to $z\sim 2$, during a period of rapid cold-gas accretion and intense in-situ SF. After this, the concentration begins to decrease. This behavior reflects the transition from early compaction phases to later inside-out growth seen in both simulations and observations of high-$z$ galaxies. Observationally, both quiescent and star-forming galaxies at $z\sim 1-3$ are known to be more compact and centrally dense than local systems at fixed mass \citep[e.g.,][]{vanderwel14,vanDokkum15}, which aligns with our results showing an increasing concentration phase, followed by its subsequent reduction. Moreover, the EBH models lie above the LBH models, basically at all times. This means that the seeding time of the SMBH plays an important role in puffing up galaxies.

Several theoretical and observational studies corroborate the interpretation that suppressed central growth produces systematically lower circular velocities. AGN-driven heating and ejective feedback due to jet-ISM coupling have been shown in observations and simulations to flatten central density profiles, reduce central baryon concentrations, and disturb the inner circular speed in Seyfert galaxies \citep{mukherjee18,garcia-burillo24,goddard25b}. Studies of halo response and core formation more generally demonstrate that energy injection from AGN can expand the inner halo and lower $V_{\rm cir}$, providing the physical mechanism that naturally explains why galaxies with suppressed central growth have lower positions on the BTFR in our suite \citep[e.g.,][]{martizzi13}. However, an alternative mechanism exists that replaces the DM cusp with a flat core due to dynamical friction \citep{elzant01,romano-diaz08}.

\subsection{Transient versus Persistent Evolutionary Signatures} 
\label{sec:time}

Understanding the physical impact of jet-mode AGN feedback in Seyfert galaxies remains challenging because the jets themselves are often unresolved or obscured at intermediate and high redshifts. In the absence of direct jet detections, galaxy scaling relations may offer an indirect diagnostic of whether galaxies have experienced sustained, intermittent, or recent AGN-driven outflows. Using results presented in this work, we can isolate which galaxy properties experience permanent, recoverable, or transient deviations due to small-scale AGN jets relative to our no-SMBH models. 

Within our simulations, we identify three distinct epochs in the evolution of the galaxy that correlate with different dynamical and feedback regimes: the early phase ($z>2.5$) dominated by major mergers; the transitional phase ($2.5>z>1$), marked by a series of minor and intermediate interactions that trigger feedback-driven regulation (see Paper\,II); and the late phase ($z<1$), where secular evolution dominates and AGN feedback continues to shape the gas and star-formation properties. Each of these epochs leaves distinct imprints on our galaxy position within the scaling relations, and we cannot interpret our results without taking into account the influences these evolutionary changes have on the properties of our galaxies. Thus, in many cases, we have referred to the evolutionary epoch or the associated environmental activity as context, in addition to the jet feedback, to aid in interpreting the evolutionary trajectory of galaxies on the scaling relations. 

The specific merger history and other environmental factors leading to the appearance of these distinct epochs do have a significant influence to the trajectory of galaxies on the scaling relations (see e.g., Figure\,\ref{fig:smbh} and related discussion). This work aims at understanding the impact of jet feedback on the parent galaxy and its environment. As such, it is an alternative to studying the statistical effects of such feedback on a population of DM halos. The latter approach is outside the scope of the present work.

In Figure\,\ref{fig:smbh}, we observe an early divergence of our AGN models from the corresponding no-SMBH fiducial models as soon as the SMBH is seeded, leaving long-lived signatures on the relation plane. All simulated galaxies dip for $\sim 1$ \,Gyr during a major merger of parent halos, ending at $z\sim3.5$. On the other hand, Figure\,\ref{fig:sfms} displays only a mild decrease in sSFR during this major merger, but AGN models experience deep dives in sSFR around $z\sim 2$, when a series of minor and intermediate mergers occur. Galaxies hosting more powerful jets show a sharp decrease in the sSFR, while galaxies with no-SMBH or with weak jets do not display this behavior. 

Overall, stronger jet and SN feedback and earlier SMBH seeding produce systematically lower stellar masses at fixed halo mass in numerical simulations. This separation persists for the remainder of their evolution.  Thus, at fixed halo mass and redshift, galaxies with lower stellar masses may preferentially host small-scale jets, with the caveat that both simulations and observations are currently unable to estimate the relative importance of jet and radiation feedback in AGN. Therefore, a consistent offset toward low $M_*$ in the $M_{\rm halo}-M_*$ relation may be an effective preliminary selection criterion for intermittent Seyfert jets, at least when analyzing simulations. The sustained nature of this divergence indicates that jet feedback regulates star formation early in the galaxy evolution. 
 
Our simulations show that specific influences on SFR can be transient, but the integrated mass deficit relative to non-AGN models grows continuously, though not monotonically. Specifically, jetted Seyferts live longer in the green valley and more frequently are moved to the quenched region in comparison to non-jetted galaxies. These transitions away from the star-forming main sequence can be short-lived, and the evolution appears to be more complex than monotonic $blue\rightarrow green\rightarrow red$.  

Galaxies at $z=0$ cluster near the median of the $M_\bullet-\sigma_{\rm bulge}$ relation, as we also observe.  Whether or not a similar relation can be expected at higher redshifts is an important and open question. Was this relation statistically true at early times, and will it continue to hold into the future?  Our galaxies do not lie on the $[M_\bullet–\sigma_{\rm bulge}]\pm 0.5$\,dex relation at $z>2$, and they approach it thereafter at steep angles. While bulges and SMBHs evolve at least partially through stochastic processes, the $M_\bullet–\sigma_{\rm bulge}$ relation does not appear to act as an attractor. From our simulations, we do not observe that the trajectories of our galaxies always return to the relation after being perturbed.  

Stronger feedback, as in the EBH and higher $\epsilon$ models, is consistently correlated with smaller $\sigma_{\rm bulge}$ and larger SMBHs (see Figure\,\ref{fig:Msigma}). This also affects the time fraction these models spend inside $\pm 0.5$\,dex from the empirical fit to the $M_\bullet-\sigma_{\rm bulge}$ relation: $\epsilon_{\rm 5,LBH} - 0\%$, $\epsilon_{\rm 15,LBH} - 54\%$, $\epsilon_{\rm 50,LBH} - 82\%$, $\epsilon_{\rm 5,EBH} - 5\%$, $\epsilon_{\rm15,EBH} - 68\%$, and $\epsilon_{\rm 50,EBH} - 95\%$.  

Jet feedback has little effect on the K-S relation in our simulations before $z\sim 2$, but after a sequence of mergers and interactions, we observe a bifurcation. The central regions of jet-hosting EBH galaxies evolve vertically downward, maintaining similar $\Sigma_{\rm H_2}$ but reduced $\Sigma_{\rm SFR}$, producing a global downward shift. By $z = 0$, these galaxies lie 1–3\,dex below their non-AGN counterparts in $\Sigma_{\rm SFR}$ at fixed gas density. We also see steeper slopes and leftward-migration on the relation during the same time frame inside $0.1R_{\rm vir}$. At $z<2$, AGN galaxies are found systematically below the K-S relation, which can be directly related to the jet activity. 

We do not see a strong sustained effect to the evolution of our galaxies on the MZR due to the presence of AGN, despite temporary fluctuations, all models trend back toward the observational median. However, in Paper\,I we do show that the metallicity gradient in the galaxy is affected by the jet feedback, becoming flatter with increasing AGN feedback strength, and in Figure\,\ref{fig:aveZ} and Paper II we show that the average CGM metallicity increases due to the presence of the AGN, and scales with the feedback strength and SMBH seeding time, i.e., in EBH. This conclusion also holds for the ISM at lower $z$, where the EBH models are more metal-rich than the LBH models. Earlier SMBH seeding, and stronger SN feedback in the EBH models, influences enrichment of the CGM and thus the metallicity of the in-falling gas at $z<0.5$. However, transient effects due to gas expulsion and accretion dominate the variation in gas-phase metallicity, especially after $z\sim 2$.

In Paper\,II we provide a detailed discussion of the formation, propagation, and effects of the jet-triggered shocks in our simulations. Here we only mention the shocks as context for the evolutionary changes to the galaxy that are measured by the scaling relations.  However, these shocks leave both transient and lasting signatures in the CGM that could be measurable by observations. The most sustained result is the enhanced CGM metallicity, as discussed already. However, the shocks themselves persist for Gyr timescales and overlap, such that they are present for basically the full (post SMBH seeding) evolution in all but one of our AGN models. The presence of these shocks around local or high-$z$ galaxies provides a clear tracer for jet activity and the specific measurable signatures are discussed further in Section\,\ref{sec:observation}.

Finally, the AGN jet feedback produces an early and lasting reduction in baryonic concentration, visible through higher $V_{\rm flat}/V_{\rm max}$ ratios and a position on the BTFR shifted toward the lower left. This signature strengthens with an earlier SMBH seeding and higher jet power and remains detectable at $z = 0$.  Similarly, we see a sustained decrease in the central stellar concentration due to the jet feedback, manifesting as larger half-mass radii compared to the non-AGN runs (see Figure\,\ref{fig:Rcomp}). Affecting the stellar mass, this divergence appears immediately upon SMBH seeding and generally grows with time, though mergers and shock-driven outflows modulate the offset magnitude differently in each model. Therefore, a galaxy with an unusually flat rotation curve, or one lying at the low-concentration end of the BTFR at fixed mass, may have experienced significant jet-driven redistribution of baryons.

In summary, across all scaling relations, a consistent picture emerges. Jet feedback in our simulated Seyfert galaxies produces a mixture of transient effects and long-lived global signatures. Several persistent deviations offer possible indirect pathways to identify jet activity, especially when the jets themselves are unresolved. These include e.g., a low stellar mass for halo mass ($M_*$ deficits), large $R_{1/2}$  or low central concentration, flatter rotation curves (i.e., higher $V_{\rm flat}/V_{\rm max}$), enriched CGM metal content, flatter metallicity gradients, and a downward offset from the K-S relation. Among the shorter-lived and less reliable indicators we identified are the instantaneous SFR, including temporarily quenched phases, and strong metallicity fluctuations due to outflows.  

\subsection{Insights and Guidance for Future Observations} 
\label{sec:observation}

Our simulations show how even low-power jets can regulate star formation, redistribute metals, and shape galactic gas distribution, these are all effects that imprint themselves on the observed galaxy scaling relations. Results presented in this work demonstrate that AGN feedback is not only critical for reproducing the quenched fractions of massive galaxies, but may also be required to match the trends observed in lower mass Seyfert-type systems. 

Importantly, while for some scaling relations, e.g., $M_{\rm halo}-M_*$, galaxies move closer to the empirical median under jet feedback, for others, e.g., K-S, it leads to offsets. These deviations away from the median may contribute to scaling relation-based diagnostics, offering a complementary pathway for detecting AGN activity that can escape traditional methods. Because these structural, kinematic, and chemical signatures persist, and often grow over cosmic time, they probe time-integrated feedback effects, rather than signatures of instantaneous processes.  

In the previous section, we have highlighted several specific signatures which, when used together, may establish a practical framework for identifying galaxies influenced by Seyfert-type jets even when the AGN is undetectable due to periods of inactivity, obscuration, sensitivity limits, or angular resolution. These include systems that (1) sit low on the $M_{\rm halo}-M_*$ relation, (2) exhibit unusually flat rotation curves or extended stellar distributions leading to a position shifted to the lower left on the BTFR, (3) display unusually metal-rich CGM halos, (4) sit low on the molecular K-S relation, possibly despite significant H$_2$ reservoirs, and (5) the detection of expanding shocks inside or even beyond the CGM. These features identified in combination provide a window into past jet activity because they persist on Gyr timescales. 

These long-term indicators may be accessible using current and upcoming facilities. These include $M_*$ deficits relative to halo mass, detectable through rest-optical/NIR SED fitting using e.g., JWST \citep{gardner06}, Euclid \citep{euclid25}, or Roman \citep{kruk25}] combined with halo mass tracers. Structural signatures, such as large $R_{1/2}$, extended stellar morphologies, and low central concentrations, are detectable in JWST imaging. Kinematic flattening is measurable using ALMA \citep{wootten09} CO at $z>1$ or JWST IFU H$\alpha$/[O\,III] at $1<z<6$. CGM metal enrichment may be identifiable through absorption-line spectroscopy with HST \citep{jakobsen89J} at low $z$ and JWST or ground-based spectrographs at higher $z$. Suppressed $\Sigma_{\rm SFR}$ at normal gas surface densities can be identified using ALMA molecular gas maps combined with JWST SFR tracers. Jet-driven shocks in our simulations have temperatures of $\sim 10^{6-7}$\,K and velocities up to $\sim 10^3\,{\rm km\,s^{-1}}$. Shocks such as these are expected to produce soft X-ray thermal emission (O\,VII/O\,VIII/Ne\,IX lines and bremsstrahlung), high-excitation UV/optical lines ([O\,III], He\,II, [N\,II], [S\,II]) and strong near-/mid-IR shock tracers ([Fe\,II], warm H$_2$, [Ne\,II], [Ne\,III], [O\,IV]) with broad, likely asymmetric profiles \citep{allen08}. These components should appear as X-ray and optical/IR bubbles or arcs. Combined gas–star–CGM diagnostics will be especially powerful as galaxies that simultaneously exhibit $M_{\rm halo}-M_*$ deficits, CGM metal enhancement, shock signatures, and K-S suppression are prime candidates for obscured or dormant jet activity.

In summary, this work provides a pathway for identifying the influence of Seyfert-like jets across cosmic time by showing that their most enduring signatures might be imprinted, not in instantaneous accretion tracers, but in the long-term structural, dynamical, and chemical evolution of their host galaxies. By following a single galaxy through multiple feedback realizations, we demonstrate which Seyfert jet-driven effects are transient and easily erased by environment, and which accumulate into lasting, measurable offsets on key scaling relations. While increasing feedback strength brings Seyfert galaxies closer to the median, their actual distribution around the scaling relations will be determined by upcoming observations. 


\begin{acknowledgments}
We thank Phil Hopkins for providing us with the latest version of the code, and are grateful to Alessandro Lupi, Da Bi, Kung-Yi Su, Paul Torrey, Xingchen Li, and Clayton Heller for their help with GIZMO. E.R.D. acknowledges support of the Collaborative Research Center 1601 (SFB 1601 sub-project C5), funded by the Deutsche Forschungsgemeinschaft (DFG, German Research Foundation) – 500700252. This work used Expanse CPU at San Diego Supercomputer Center (SDSC) through allocation PHY230135 from the ACCESS program, which is supported by NSF grants 2138259, 2138286, 2138307, 2137603, and 2138296 \citep{boerner23}, and by the University of Kentucky Morgan Computing Cluster. We are grateful for help by Vikram Gazula at the Center for Computational Studies of the University of Kentucky.
\end{acknowledgments}

\section*{Data Availability}
 The data used for this paper will be shared upon any reasonable request which intends the data to be used for scientific endeavors. Requests can be made by email to the authors. The available data includes figures, ascii tables of galaxy and CGM data, and simulation snapshot files. The method of data transfer will be determined on a case by case basis as we will need to determine an appropriate method based on the specific request.

\bibliography{paper}{}
\bibliographystyle{aasjournalv7}



\end{document}